\documentclass[useAMS, usenatbib, usegraphicx]{mn2e} 
\usepackage{amssymb}

\newcommand{\cMpch}{~h^{-1}~\mbox{comoving Mpc}}
\newcommand{\Mpch}{~h^{-1}~\mbox{Mpc}}

\newcommand{\invMpc}{~\mbox{Mpc}^{-3}}
\newcommand{\yr}{~\mbox{yr}}
\newcommand{\invyr}{~\mbox{yr}^{-1}}
\newcommand{\invs}{~\mbox{s}^{-1}}
\newcommand{\Msunh}{~h^{-1}~\mbox{M}_{\odot}}
\newcommand{\Msun}{~\mbox{M}_{\odot}}
\newcommand{\K}{~\mbox{K}}
\newcommand{\cmci}{~\mbox{cm}^{-3}}
\newcommand{\cmsi}{~\mbox{cm}^{-2}}
\newcommand{\eV}{~\mbox{eV}}
\newcommand{\kms}{~\mbox{km s}^{-1}}
\newcommand{\gadget}{{\sc gadget-2}}
\newcommand{\cmbfast}{{\sc cmbfast}}
\newcommand{\cloudy}{{\sc cloudy}}

\title[Keeping the Universe ionised]{Keeping the Universe ionised:
  Photo-heating and the clumping factor of the high-redshift intergalactic medium}

\author[Pawlik, Schaye \& van Scherpenzeel] 
       {Andreas H. Pawlik$^{1}$\thanks{E-mail: pawlik@strw.leidenuniv.nl},
	 Joop Schaye$^{1}$\thanks{E-mail: schaye@strw.leidenuniv.nl}
	 and 
	 Eveline van Scherpenzeel$^{1}$\\
	 $^{1}$Leiden Observatory, Leiden University, P.O. Box 9513, 2300RA Leiden, The Netherlands}

\begin{document}

\date{Accepted; Received; in original form}

 \pagerange{\pageref{firstpage}--\pageref{lastpage}} \pubyear{}

\maketitle

\label{firstpage}

\begin{abstract}
The critical star formation rate density required to keep the intergalactic hydrogen 
ionised depends crucially on the average rate of recombinations in the 
intergalactic medium (IGM). This rate is proportional to the 
clumping factor $C \equiv \langle \rho_{\rm{b}}^2 \rangle_{\rm IGM} / \langle \rho_{\rm{b}} \rangle^2$, where
$\rho_{\rm{b}}$ and $\langle \rho_{\rm{b}} \rangle$ are the local and cosmic mean baryon density, respectively 
and the brackets  $\langle \rangle_{\rm IGM}$ indicate spatial averaging over the recombining gas in the IGM. 
We perform a suite of cosmological 
smoothed particle hydrodynamics simulations that include radiative cooling to calculate the volume-weighted clumping factor of the IGM at 
redshifts $z \ge 6$. We focus on the effect of photo-ionisation 
heating by a uniform ultra-violet background and find that photo-heating strongly reduces the clumping factor
because the increased pressure support smoothes out small-scale density fluctuations. 
Photo-ionisation heating is often said to provide a negative feedback on the reionisation of the IGM because it suppresses the cosmic star formation
rate by boiling the gas out of low-mass halos. However, because of the reduction of the clumping factor
it also makes it easier to keep the IGM ionised. Photo-heating therefore also provides a positive feedback which, 
while known to exist, has received much less attention. We demonstrate that this positive feedback is in fact very strong.
Using conservative assumptions, we find that if the IGM was reheated at $z \gtrsim 9$, 
the observed population of star-forming galaxies at $z \approx 6$ may be sufficient
to keep the IGM ionised, provided that the fraction of ionising photons that escape the star-forming regions to ionise the IGM is
larger than $\sim 0.2$. 
\end{abstract}


\section{Introduction}
The absence of a Gunn-Peterson trough in the majority of the observed absorption spectra towards 
high-redshift quasars suggests that the reionisation of intergalactic hydrogen was  
completed at a redshift $z \gtrsim 6$ (see \citealp{Fan:2006} for a recent review) and that it remained 
highly ionised ever since. Current observational estimates of the ultra-violet (UV) luminosity 
density at redshifts $z \lesssim 6$ 
(e.g.~\citealp{Stanway:2003}; \citealp{Lehnert:2003}; 
\citealp{Bunker:2004}; \citealp{Bouwens:2004}; \citealp{Yan:2004}; \citealp{Sawicki:2006}; \citealp{Bouwens:2006};
\citealp{Mannucci:2007}; \citealp{Oesch:2008}; \citealp{Bouwens:2008}), 
on the other hand, may imply star formation rate (SFR) densities several times lower than the critical SFR density
required to keep the intergalactic medium (IGM) ionised (but see, e.g.,
\citealp{Stiavelli:2004}; \citealp{Malhotra:2005}; \citealp{Panagia:2005}). 
Taken at face value, these low SFR densities pose a severe challenge to commonly employed 
theoretical models in which the observed population of star-forming 
galaxies is the only source of ionising radiation in the high-redshift Universe.
\par
There are, however, large uncertainties associated
with both the observationally inferred (see, e.g., the comprehensive analysis of ~\citealp{Bouwens:2007}) 
and the critical SFR densities. The critical SFR density,
\begin{eqnarray}
  \dot{\rho}_* &\approx&  0.027 \Msun \invyr \invMpc \nonumber \\
  &\quad& \times {f_{\rm{esc}}}^{-1} \left ( \frac{C}{30} \right )  \left ( \frac{1+z}{7} \right )^3  \left ( \frac{\Omega_{\rm{b}}h_{70}^2}{0.0465} \right )^2,
  \label{Eq:CriticalSfr}
\end{eqnarray}
here rescaled to match the most recent WMAP estimate for the cosmic baryon density (\citealp{Komatsu:2008}),
has been derived by \cite{Madau:1999} using an early version of the \cite{Bruzual:2003} population synthesis code, assuming a 
Salpeter initial stellar mass function (IMF) and solar metallicity. It results
from simply equating the spatially averaged rate at which ionising photons are emitted into the IGM 
to the spatially averaged rate at which the intergalactic gas recombines. Eq.~\ref{Eq:CriticalSfr} is therefore incapable of
addressing a number of potentially important physical effects. Some ionising photons will, for instance,
redshift below the ionisation threshold before ionising and some
ionising photons will have been emitted longer than a recombination time
ago upon impact with a neutral atom, so that equating instantaneous rates is
not appropriate and one even may have to take source evolution into account.
It is therefore important to keep in mind that Eq.~\ref{Eq:CriticalSfr} is likely only accurate within factors of a few.
\par
The critical SFR is inversely proportional to the escape fraction $f_{\rm{esc}}$, i.e. 
the fraction of ionising photons produced by star-forming galaxies
that escape the interstellar medium (ISM) to ionise the IGM, 
and proportional to the average recombination rate in the IGM. 
The latter is parametrised 
using the dimensionless clumping factor $C \equiv \langle \rho_{\rm{b}}^2\rangle_{\rm IGM}  / \langle \rho_{\rm{b}}\rangle^2$, 
where $\rho_{\rm{b}}$ is the baryon density, $\langle \rho_{\rm{b}}\rangle$ is the mean baryon 
density of the Universe and the brackets $\langle \rangle_{\rm IGM}$ indicate spatial averaging over the gas constituting the recombining IGM.
Under the assumption of a uniformly ionised IGM, 
the clumping factor expresses the spatially averaged number of 
recombinations occurring per unit time and unit volume in the ionised IGM, relative to that
in gas at the cosmic mean density $\langle \rho_{\rm{b}}\rangle$. 
A larger escape fraction implies a smaller critical SFR density, as more photons are
available to ionise the IGM. On the other hand, a larger clumping factor 
implies a larger critical SFR density since more ionising photons are required to compensate for 
the increased number of recombinations.
\par
Most observational studies that compare\footnote{Note that although 
  the critical SFR density is sensitive to the IMF, this comparison is insensitive to the IMF provided the same IMF is used to compute
the critical and observationally derived SFR densities. This is because the UV luminosity density is dominated by the same 
massive stars that are responsible for the emission of ionising photons with energies $>13.6 \eV$ (\protect\citealp{Madau:1999}).} 
the SFR density derived from estimates of the UV luminosity density
at redshift $z \approx 6$ to the critical SFR density assume an escape fraction $f_{\rm esc}
\lesssim 0.5 $ and a clumping factor $C = 30$. While 
a variety of both observational
and theoretical studies (e.g. \citealp{Inoue:2006} and references therein; \citealp{Razoumov:2006}; \citealp{Gnedin:2008a})
have ruled out larger escape fractions, the estimate for the clumping factor
comes from a single cosmological simulation performed
more than 10 years ago (\citealp{Gnedin:1997}). 
It is on the basis of these values for the escape fraction and the clumping factor 
that the observed population of galaxies has 
been found to be incapable of keeping the intergalactic hydrogen ionised, forming massive stars
at a rate which is up to an order of magnitude lower than required by Eq.~\ref{Eq:CriticalSfr}. 
\par
It has been pointed out that this discrepancy between the inferred and critical SFR densities could be resolved
if the employed clumping factor were too high (e.g.~\citealp{Sawicki:2006}; 
see also the discussion in \citealp{Bouwens:2007}). Indeed, in most (but not all) of the more recent theoretical theoretical studies
(e.g.~\citealp{Valageas:1999}; \citealp{Miralda:2000}; \citealp{Gnedin:2000a}; \citealp{Haiman:2001}; \citealp{Benson:2001}; \citealp{Chiu:2003};
\citealp{Iliev:2007}; \citealp{Srbinovsky:2007}; \citealp{Kohler:2007}; \citealp{Bolton:2007}; \citealp{Maio:2007}; \citealp{Furlanetto:2008})
significantly lower clumping factors were derived. On the other hand, it is sometimes emphasised that simulations 
underestimate the clumping factor, due to a lack of resolution (see, e.g., \citealp{Madau:1999}). 
In this work we perform a set of cosmological Smoothed Particle Hydrodynamics (SPH) simulations 
that include radiative cooling and photo-ionisation by a uniform UV background in 
the optically thin limit to study the clumping factor of the IGM.
\par
We focus on the effect of photo-ionisation heating on the evolution of the clumping factor.
Previous investigations of the impact of photo-heating on the reionisation of the IGM
have almost exclusively come to the conclusion that it acts as to provide a negative feedback. 
Photo-heating boils the gas out of the potential wells of dark matter (DM) halos with virial temperatures 
$T_{\rm vir} \lesssim 10^4\K$ (e.g.~\citealp{Thoul:1996}; \citealp{Navarro:1997};
\citealp{Barkana:1999}; \citealp{Kitayama:2000}; \citealp{Gnedin:2000b}; \citealp{Dijkstra:2004};
\citealp{Shapiro:2004}; \citealp{Hoeft:2006}; \citealp{Crain:2007}; \citealp{Mesinger:2008}; \citealp{Okamoto:2008}; \citealp{Pawlik:2008}). 
This inhibits the formation of stars in these low-mass halos and thus decreases the ionising emissivity, which makes it more difficult 
to reionise the Universe. The same mechanism that reduces the number of ionising photons that are emitted into the IGM 
does, however, also affect the evolution of the clumping factor 
(e.g.~\citealp{Haiman:2001}; \citealp{Oh:2003}; \citealp{Kuhlen:2005}; \citealp{Wise:2005}; \citealp{Furlanetto:2006}; \citealp{Ciardi:2007}).
\par
In this paper we demonstrate that photo-heating significantly lowers the clumping factor and hence the average recombination rate in the 
IGM.  While photo-ionisation heating undoubtedly impedes the production of ionising photons, our results imply 
that it also makes it much easier to keep the IGM ionised. 
\par
The paper is structured as follows. In Section~\ref{Sec:Simulations}
we give a detailed description of our set of simulations. In Section~\ref{Sec:Results}
we use our simulations to compute the clumping factor of the IGM. Finally, in Section~\ref{Sec:Conclusions}, 
we discuss our results and their implications for the value of the critical SFR density. 

\section{Simulations}
\label{Sec:Simulations}
\begin{table*}
\begin{center}
\caption{Simulation parameters:
  comoving size of the simulation box, $L_{\rm box}$ (default value: $6.25 \Mpch$);
  number of DM particles, $N_{\rm dm}$ (default value: $256^3$);
  mass of dark matter particles, $m_{\rm dm}$ (default value: $8.6 \times 10^5\Msunh$);
  additional reheating energy per proton, $\epsilon_{\rm r}$ (default value: $2 \eV$);
  reheating  redshift, $z_{\rm{r}}$ (default value: $9$);
  kinetic feedback from supernova winds, winds (default: no);
  cosmological parameters, WMAP (default: $5$-year).
  The number of SPH particles initially equals $N_{\rm dm}$ (it decreases
  during the simulation due to star formation). Bold font indicates our default simulation.
\label{tbl:params}}
\begin{tabular}{ccccccccc}

\hline
\hline
Simulation & $L_{\rm box}$ & $N_{\rm dm}$ & $m_{\rm dm}$  & $\epsilon_{\rm r}$ &  $z_{\rm{r}} $ & winds& WMAP &\\
           & $[\cMpch]$   &             & $[10^5\Msunh]$  & $[\eV]$          &   &     &            year& \\           
\hline
\bf \textit{r9L6N256}        &  $\bf 6.25$ &  $\bf 256^3$ &  $\bf 8.6$ &  $\bf 2$ & $\bf 9$ & \bf no &  $\bf 5$&\\
\textit{L6N256}        & $6.25$ & $256^3$ & $8.6 $ & $0$ & $0$ & no & $5$&\\
\textit{r[$z_{\rm{r}}$]L6N256}   & $6.25$ & $256^3$ & $8.6$ & $2$ & $[7.5, 10.5, 12, 13.5, 15, 19.5]$ &  no & $5$&\\
\textit{r9L6N256highT}   & $6.25$ & $256^3$ & $8.6$ & $20$ & $9$ &  no & $5$& \\
\textit{r9L6N256lowT}   & $6.25$ & $256^3$ & $8.6$ & $0$ & $9$ & no & $5$&\\
\textit{r9L6N256winds}        & $6.25$ & $256^3$ & $8.6$ & $2 $ & $9$ & yes & $5$&\\
\textit{r9L6N256W1}        & $6.25$ & $256^3$ & $8.3$ & $2$ & $9$ & no & $1$&\\
\textit{r9L6N256W3}        & $6.25$ & $256^3$ & $7.9$ & $2 $ & $9$ & no & $3$&\\
\textit{r9L12N256}        & $12.5$ & $256^3$ & $69.1$ & $2 $ & $9$ & no & $5$&\\
\textit{r9L6N128}        & $6.25$ & $128^3$ & $69.1$ & $2$ & $9$ & no & $5$&\\
\textit{r9L6N064}        & $6.25$ & $64^3$ & $552.8$ & $2$ & $9$ & no & $5$&\\
\textit{r9L3N064}         & $3.125$ & $64^3$ & $69.1$ & $2 $ & $9$ &  no & $5$&\\

\hline
\end{tabular}
\end{center}
\end{table*}

\begin{figure}
  \includegraphics[width=0.49\textwidth]{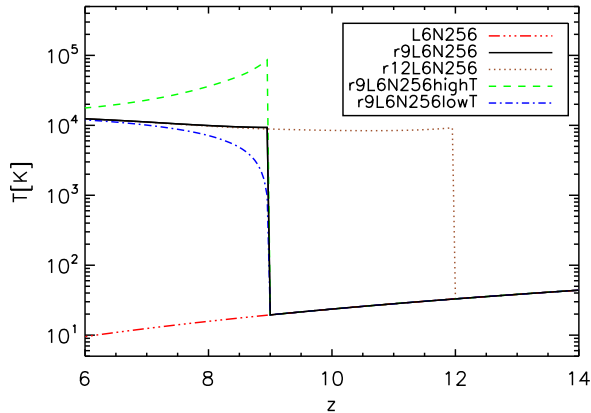}
  \caption{Thermal evolution of gas with overdensity $\Delta = 1$ for characteristic choices of the 
    reheating parameters $z_{\rm{r}}$ and $\epsilon_{\rm{r}}$, which   are listed in Table~\ref{tbl:params} 
    for the simulations indicated in the legend. Note that even in the absence
    of an additional energy input at redshift $z = z_{\rm{r}}$, i.e. for  $\epsilon_{\rm r} = 0$,  as it is the case 
    for simulation \textit{r9L6N256lowT}, the gas is quickly heated by the UV background to a temperature $T \sim 10^4 \K$.}
  \label{Fig:ThermEvol}
\end{figure}
We use a modified version of the N-body/TreePM/SPH code \gadget\ 
(\citealp{Springel:2005}) to perform a suite of cosmological SPH simulations including radiative cooling.
\par
The initial particle positions and velocities are obtained from glass-like initial conditions
using \cmbfast\ (version 4.1; \citealp{Seljak:1996}) and employing the Zeldovich approximation
to linearly evolve the particles down to redshift z = 127. We assume a flat $\Lambda$CDM universe
and employ the set of cosmological parameters 
$[\Omega_{\rm{m}}, \Omega_{\rm{b}}, \Omega_\Lambda, \sigma_8, n_{\rm{s}}, h]$ given by
$[0.258, 0.0441, 0.742, 0.796, 0.963, 0.719]$,
in agreement with the WMAP 5-year observations (\citealp{Komatsu:2008}). For comparison, we also perform some simulations
employing the set of cosmological parameters $[0.238, 0.0418, 0.762, 0.74, 0.951, 0.73]$ and
$[0.25, 0.045, 0.75, 0.9, 1, 0.73]$, consistent with WMAP 3-year (\citealp{Spergel:2007}) and
WMAP 1-year (\citealp{Spergel:2003}) observations, respectively.
Data is generated at $50$ equally spaced redshifts between $z = 20$ and $z = 6$. 
The parameters of the simulations employed for the present work are summarised in Table~\ref{tbl:params}.
\par
The gravitational forces are softened over a length of $1/25$ of the mean dark matter 
inter-particle distance. We employ the star formation recipe of \cite{Schaye:2008}, to which we refer the
reader for details. Briefly, gas with densities exceeding the
critical density for the onset of the thermo-gravitational
instability (hydrogen number densities $n_{\rm{H}} = 10^{-2} - 10^{-1}\cmci$) is expected to be
multiphase and star-forming (\citealp{Schaye:2004}). We therefore
impose an effective equation of state (EoS) with pressure
$P\propto \rho^{\gamma_{\rm{eff}}}$ for densities $n_{\rm{H}} > n_{\rm{H}}^*$, where $n_{\rm{H}}^* \equiv 10^{-1}\cmci $, normalised
to $P/k = 10^3 \cmci \K$ at the critical density $n_{\rm{H}}^*$. We use $\gamma_{\rm{eff}} = 4/3$
for which both the Jeans mass and the ratio of the Jeans
length and the SPH kernel are independent of the density,
thus preventing spurious fragmentation due to a lack of
numerical resolution. Gas on the effective EoS is allowed to form stars using a pressure-dependent rate that reproduces
the observed Kennicutt-Schmidt law (\citealp{Kennicutt:1998}), renormalised by a 
factor\footnote{This conversion factor between SFRs 
has been computed using the \cite{Bruzual:2003} population synthesis code
for model galaxies of age $> 10^7 \yr$ forming stars at a constant rate and 
is insensitive to the assumed metallicity.} of $1/1.65$ to account for the fact that it assumes a 
Salpeter IMF whereas we are using a Chabrier IMF. 
\par
The gas is of primordial composition, with a hydrogen mass fraction $X = 0.752$ and a helium mass fraction $Y = 1-X$.
Radiative cooling and heating are included assuming ionisation equilibrium, 
using tables generated with the publicly available package \cloudy\ (version 05.07 of the code last described by \citealp{Ferland:1998}),
as described in \cite{Wiersma:2008}.
The gas is allowed to cool by collisional ionisation and excitation, 
emission of free-free and recombination radiation and Compton cooling off the cosmic microwave background.
\par
We perform a set of simulations including photo-ionisation by a uniform UV background in the optically thin limit  
at redshifts below the reheating redshift $z_{\rm{r}}$. These simulations are denoted with the 
prefix \textit{r} (see Table~\ref{tbl:params}). To study the effect of reionisation reheating,
we compare these simulations to a simulation that does not include photo-ionisation (\textit{L6N256}).
Note that the photo-ionisation changes the density of free electrons and the ionic abundances. 
Both the cooling and heating rates are therefore affected by the inclusion of
a UV background (e.g.~\citealp{Efstathiou:1992}; \citealp{Wiersma:2008}).
\par
The properties of the UV background depend on the redshift of reheating.
If $z_{\rm{r}} \le 9$, we employ the evolving UV background from quasars and
galaxies tabulated by \cite{Haardt:2001} for $z \le z_{\rm{r}}$. If $z_{\rm{r}} > 9$, we use 
the $z = 9$ \cite{Haardt:2001} UV background for all redshifts $ 9 < z \le z_{\rm{r}}$, and employ the evolving \cite{Haardt:2001} 
UV background for redshifts $z \le 9$. This is necessary because \cite{Haardt:2001} only tabulate up to $z = 9$. 
For $z > z_{\rm{r}}$, we employ the  $z = 9$ \cite{Haardt:2001} UV background 
but with its intensity at energies equal to and larger than $13.6 \eV$ set to zero. 
Molecular hydrogen and deuterium and their catalysts are kept photo-dissociated by this soft UV background at all redshifts 
and therefore never contribute to the cooling rate. Our approach is motivated in the context of reionisation 
because the weak UV background established by the very first ionising sources is already sufficient to 
efficiently suppress the formation of molecular hydrogen 
(e.g.~\citealp{Haiman:1997} and references therein; \citealp{Glover:2007}; \citealp{Chuzhoy:2007}). 
\begin{figure*}
  \includegraphics[width=0.31\textwidth]{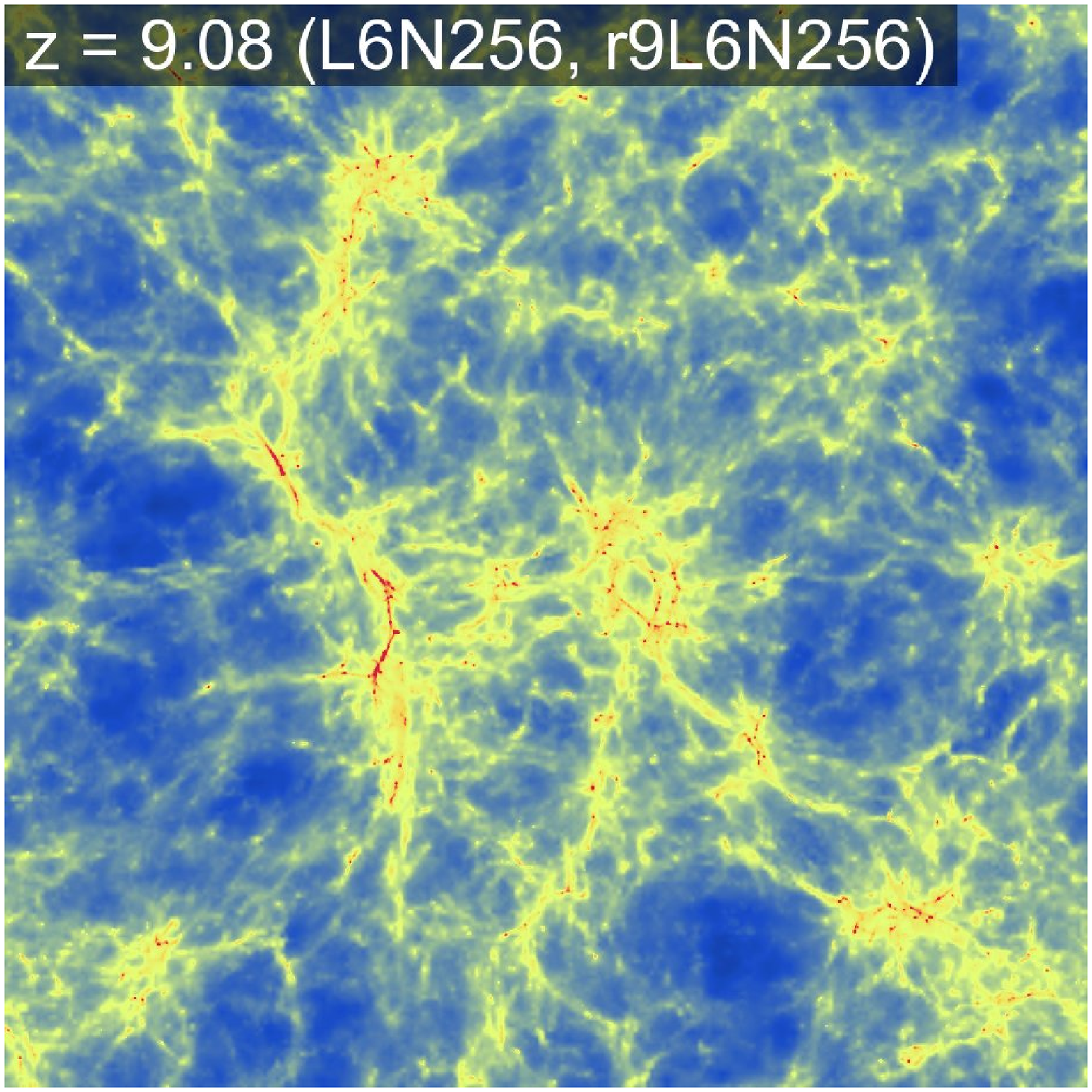}
  \includegraphics[width=0.31\textwidth]{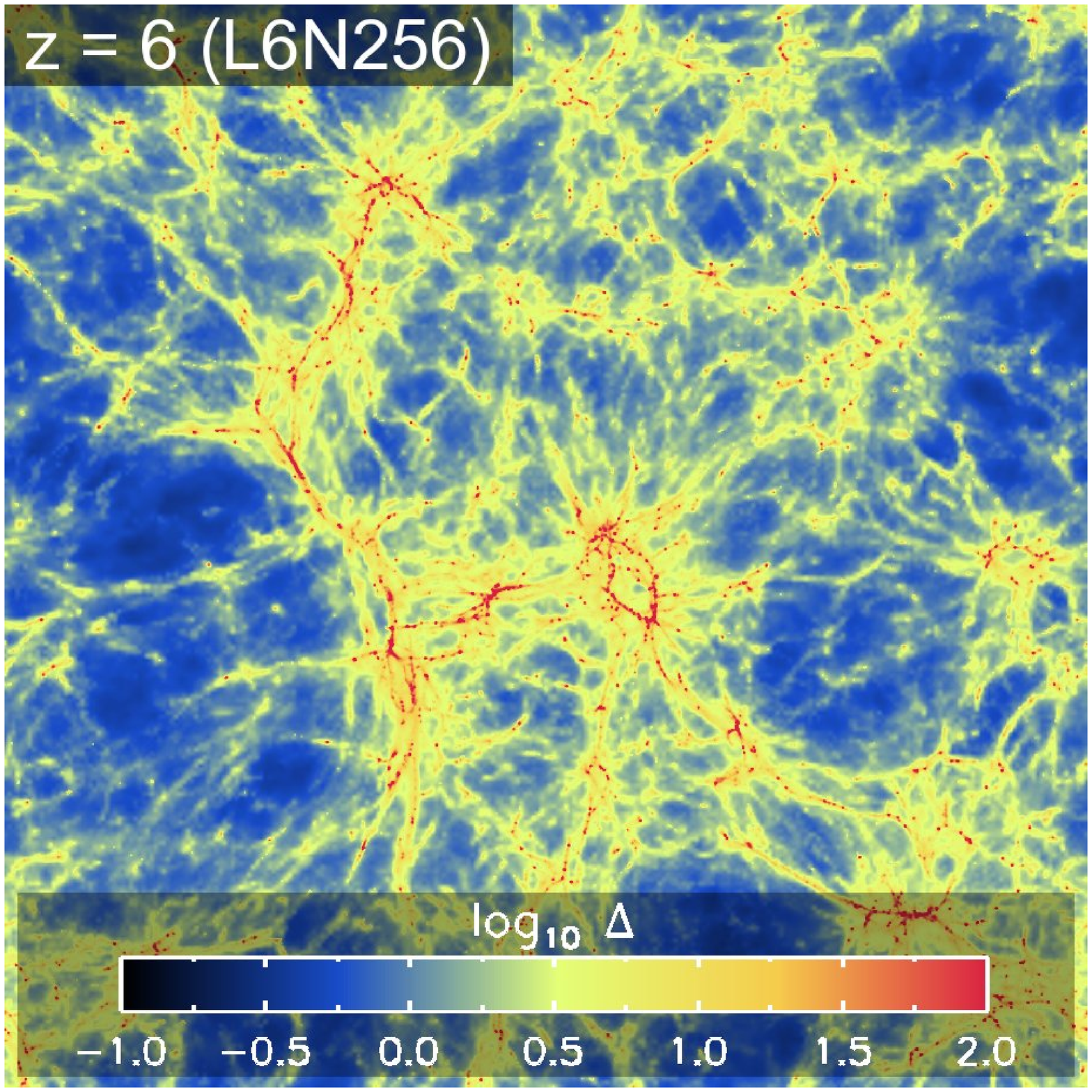}
  \includegraphics[width=0.31\textwidth]{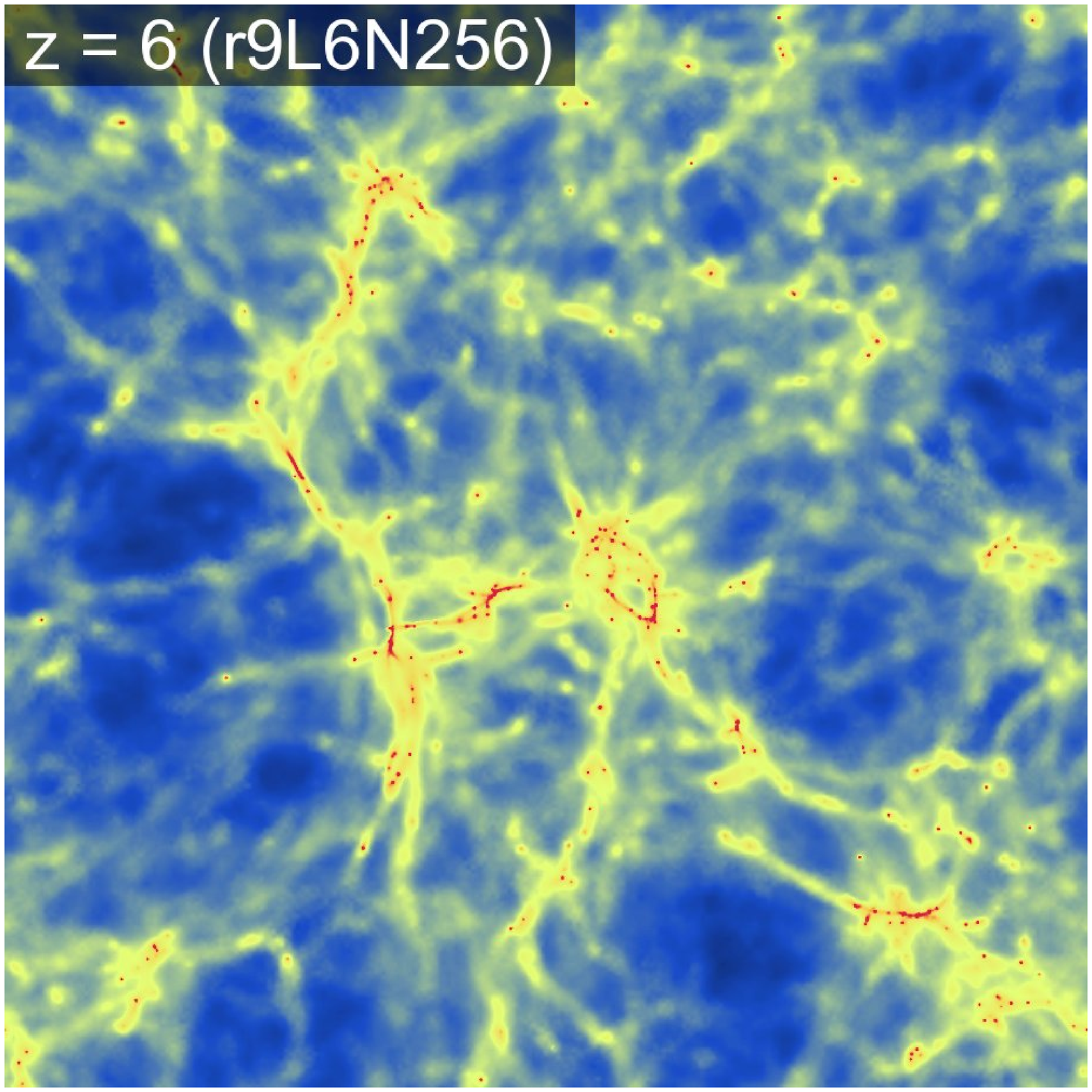}
  \caption{Slices (of thickness $1.25 \cMpch$) through the centre of the simulation box, showing the SPH overdensity field  
    in the simulations \textit{L6N256} and \textit{r9L6N256} at redshifts $z = 9.08$ (left-hand panel; where they are identical) and $z = 6$
    (middle panel: \textit{L6N256}, right-hand panel: \textit{r9L6N256}). The inclusion of photo-heating in \textit{r9L6N256} 
    leads to a strong smoothing of the density field (right-hand panel).}
  \label{Fig:Slices}
\end{figure*}
\par
The reheating redshift $z_{\rm r}$  is a parameter in our simulations. The most recent determination of the
Thomson optical depth towards reionisation from the WMAP (5-year) experiment implies a reionisation redshift 
$z_{\rm reion} = 11.0 \pm 1.4$, assuming that the transition from the neutral to the fully ionised Universe was instantaneous
(\citealp{Komatsu:2008}).
The Thomson optical depth towards reionisation provides, however, only an integral constraint on the Epoch of Reionisation.
The reionisation history may therefore have been considerably more intricate. 
An early population of X-ray sources, for example, could reheat the IGM to temperatures $\sim 10^4 \K$
already at much higher redshifts (e.g.~\citealp{Collin:1991}; \citealp{Madau:1999a}; \citealp{Oh:2001}; 
\citealp{Venkatesan:2001}; \citealp{Machacek:2003}; \citealp{Madau:2004}; \citealp{Ricotti:2004}).
We therefore study a range of thermal histories, performing simulations using
$z_{\rm r} = 7.5, 9, 10.5, 12, 13.5, 15$ and $19.5$.
To be conservative, we use the relatively low reheating redshift $z_{\rm r} = 9$ as our
default value.
\par
In our simulations we compute the photo-heating rates in the optically thin limit, which means that we underestimate 
the temperature of the IGM during reionisation (e.g.~\citealp{Abel:1999}). 
We therefore inject an additional thermal energy $\epsilon_{\rm{r}}$ per proton
at $z = z_{\rm{r}}$ (see, e.g.,  \citealp{Thoul:1996}). By varying the parameter
$\epsilon_{\rm{r}}$, we will investigate the sensitivity of our
results to the temperature of the reheated IGM. Our default simulation employs $\epsilon_{\rm r} = 2 \eV$.
Fig.~\ref{Fig:ThermEvol} shows the thermal evolution of gas at the cosmic mean baryon density $\langle \rho_{\rm{b}}\rangle$, 
i.e. of gas with overdensity $\Delta \equiv \rho_{\rm{b}}/\langle \rho_{\rm{b}}\rangle = 1$, 
for different values of $\epsilon_{\rm{r}}$ and $z_{\rm{r}}$. At $z = z_{\rm r}$, the gas is heated to
$T_{\rm r} \approx 10^4 \K$ for $\epsilon_{\rm r} = 2 \eV$, whereas the gas temperature is about an order of magnitude higher (lower) for
$\epsilon_{\rm r} = 20 \eV$ ($\epsilon_{\rm r} = 0 \eV$). After reheating the gas quickly looses memory of its initial 
temperature and by $z = 6$ the gas temperature is $T \approx 10^4 \K$ in all cases.
\par
In one of our simulations (\textit{r9L6N256winds}) we include kinetic feedback
from star formation. We employ the prescription of \cite{caius:2008}, which is a
variation of the \cite{Springel:2003} recipe for kinetic feedback. In this prescription,
core-collapse supernovae locally inject kinetic energy and kick gas particles
into winds. The feedback is specified by two parameters, the mass
loading $\eta \equiv \dot{M}_w/\dot{M}_*$, which describes the initial wind
mass loading $\dot{M}_w$ in units of the cosmic SFR
$\dot{M}_*$, and the initial wind velocity $v_w$. We use $\eta = 2$ and $v_w = 600
\kms$,  consistent with observations of local (e.g.~\citealp{Veilleux:2005}) and redshift $z
\approx 3$ (\citealp{Shapley:2003}) starburst galaxies. Note that wind particles are not
hydrodynamically decoupled and that they are launched local to the star formation event, different from the \cite{Springel:2003} recipe.
\par
\section{Results}
\label{Sec:Results}
In this section we employ the set of simulations described in Section~\ref{Sec:Simulations} and summarised in 
Table~\ref{tbl:params} to calculate the clumping factor of the IGM. We start in Section~\ref{Sec:PDF} with analysing the distribution of
the gas in our default simulation \textit{r9L6N256} and in the simulation \textit{L6N256}, which is identical
to our default simulation except for the fact that it does not include a photo-ionising background.
In Section~\ref{Sec:ClumpingFactor} we discuss the definition of the clumping factor and compare the clumping factors
derived from our default simulation \textit{r9L6N256} to that derived from simulation \textit{L6N256}. 
We discuss the convergence of our results with respect to variations in the mass resolution and 
in the size of the simulation box in Section~\ref{Sec:Convergence}. 
In Section~\ref{Sec:ReheatingRedshift} we vary the redshift at which
the ionising UV background is turned on and in Section~\ref{Sec:ReheatingTemperature} we demonstrate that our 
conclusions are robust with respect to our choice for the temperature to which the IGM
is photo-heated. In Section~\ref{Sec:Feedback} we discuss 
how kinetic feedback from supernova winds affects our results and 
quote the clumping factors obtained from the simulations employing WMAP 3-year and 1-year cosmological parameters. We conclude with
a brief comparison to previous work.
\par
\begin{figure*}
  \includegraphics[width=0.49\textwidth]{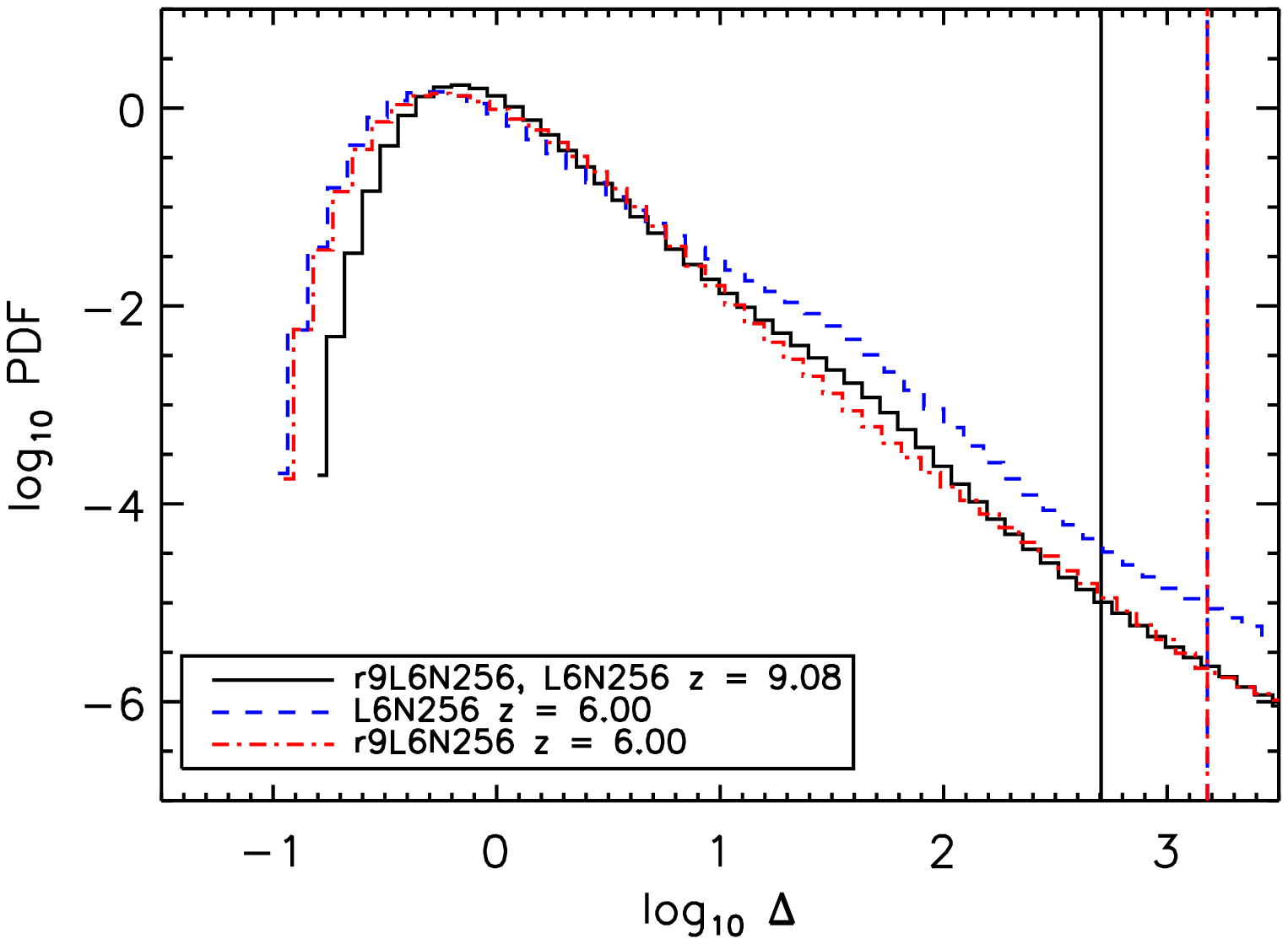}
  \includegraphics[width=0.49\textwidth]{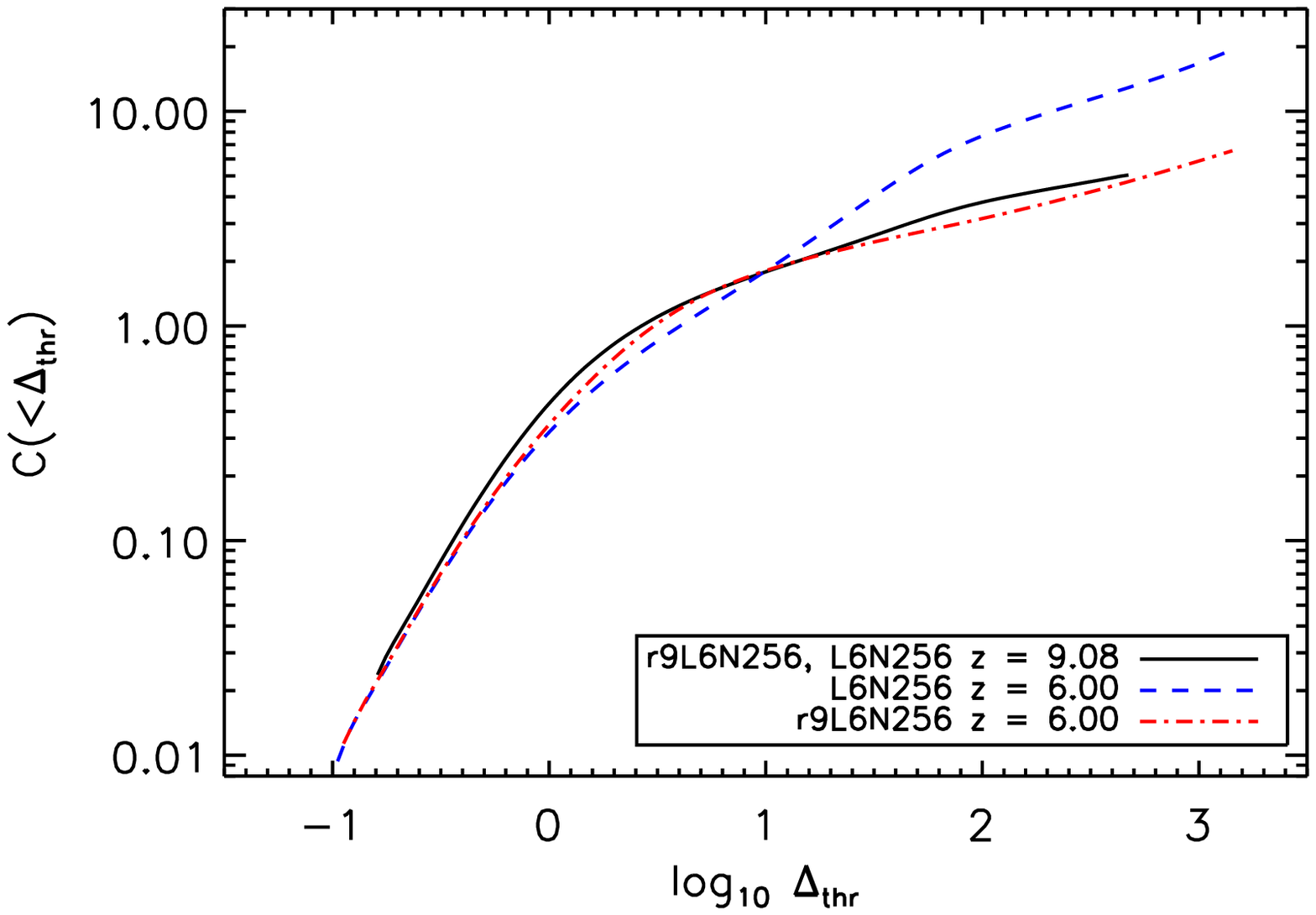}

  \caption{{\it Left-hand panel:} Volume-weighted PDF of the baryon overdensity $\Delta$ per unit $\log_{10} \Delta$ for simulations
    \textit{r9L6N256} and \textit{L6N256} at redshifts $z = 9.08$ and $z = 6$, as indicated in the legend. 
    Photo-heating destroys the bump around overdensities $1 \lesssim \log_{10}\Delta \lesssim 2$, 
    which mark the gas that accretes onto DM halos. The vertical lines
    (which match the colour and style of the corresponding PDFs)
    indicate the overdensities corresponding to the 
    onset of star formation.
    {\it Right-hand panel:} Clumping factor $C(< \Delta_{\rm thr})$ of gas with overdensity 
    $\Delta < \Delta_{\rm thr}$ for the simulations shown in the left-hand panel. The inclusion of photo-heating
    in \textit{r9L6N256} leads to a clumping factor that is substantially smaller than that obtained from \textit{L6N256},
    for threshold overdensities $\Delta_{\rm thr} > 10$. Note that the maximum threshold overdensities we consider for 
    the calculation of the clumping factor are given by the critical 
    density $n_{\rm H}^* \equiv 10^{-1} \cmci$ for the onset of star-formation (the vertical lines shown in the left-hand panel).}
  \label{Fig:Cmax}
\end{figure*}
\par
\subsection{The gas density distribution}
\label{Sec:PDF}
Here we compare the gas distributions in our default simulation \textit{r9L6N256} (in which the UV background 
is turned on at redshift $z_{\rm r} = 9$) and in the simulation \textit{L6N256} (which does not include photo-ionisation heating). 
\par
Figure~\ref{Fig:Slices} shows the overdensities at redshifts $z = 9.08$ and $z = 6$ in a slice through the simulation box  
for these simulations. Heating by the photo-ionising background almost 
instantaneously increases the gas temperatures to $T_{\rm r} \sim 10^4 \K$
(see Fig.~\ref{Fig:ThermEvol}) and accordingly raises the cosmological Jeans mass.
Gas that had already settled into the potential wells of DM halos with virial temperature $T_{\rm vir} \lesssim T_{\rm r}$ 
is driven back into the diffuse IGM by the increased pressure gradient forces (e.g.~\citealp{Barkana:1999}; \citealp{Shapiro:2004}).
The large cosmological Jeans mass prevents any re-accretion of gas or infall of previously unbound gaseous material 
into these low-mass halos and keeps the IGM diffuse. Comparing the middle panel with the right-hand panel of Fig.~\ref{Fig:Slices}, 
this {\it Jeans filtering} (e.g.~\citealp{Shapiro:1994}; \citealp{Gnedin:1998}; \citealp{Gnedin:2000b}; \citealp{Okamoto:2008}) 
in the presence of photo-heating leads to a strong smoothing of the small-scale density fluctuations by $z = 6$.
\par
A detailed analysis of the overdensity distribution in the simulations can be obtained by studying  
$\mathcal{P}_{\rm V}(\Delta)$, the volume-weighted probability density function (PDF) for $\Delta$.
We show the PDF (per unit $\log_{10} \Delta$ and normalised according to
$\int_0^{\infty} d\Delta\ \mathcal{P}_{\rm V}(\Delta) = 1$) in the left-hand panel of Fig.~\ref{Fig:Cmax}. 
It is important to be aware of the fact that the finite numerical resolution of our simulations implies an
unavoidable intrinsic smoothing of the gas density distribution on the scale of the
SPH kernel or the scale over which the gravitational forces are softened,
whichever is larger. A numerical smoothing on scales larger than the Jeans filtering scale 
(below which the gas density distribution is physically smoothed) would prevent us from obtaining converged results. 
We will discuss the convergence of our simulations with respect to resolution in Section~\ref{Sec:Convergence}.
\par
At redshift $z=9.08$ (black solid histogram), the gravitational amplification of the overdensities present in the initial conditions
has produced a significant deviation of the PDF from its primordial Gaussian shape.
The flattening of the slope of the PDF for overdensities $1 \lesssim \log_{10}\Delta \lesssim 2$
can be attributed to the shock-heating of gas falling into the potential wells
of dark matter halos, most of which have virial temperatures $\lesssim 10^4 \K$,
which we refer to as low-mass halos.  The shape of the PDF
is determined by the effective EoS once the gas reaches the critical density for the onset of star formation 
($n^*_{\rm H} \equiv 10^{-1}\cmci$, see Section~\ref{Sec:Simulations}; indicated by the vertical lines). 
\par
At redshifts $z < z_{\rm r}$, the shape of the PDF strongly depends on whether photo-heating by the ionising background is included or not.
In the absence of such a background (\textit{L6N256}, blue dashed histogram), gravitational collapse proceeds unimpeded, increasing the 
PDF at $\log_{10} \Delta \gtrsim 1$. Since the gas that accretes onto DM halos must originate from the reservoir 
at $\log_{10} \Delta \lesssim 1$ (the diffuse IGM), the PDF decreases over this range of overdensities. 
As a result, the maximum of the PDF shifts to lower overdensities.
At the same time, the gravitational amplification of large-scale underdense regions 
leads to an increase in the PDF around overdensities $\log_{10} \Delta \sim -1$.
\par
Photo-heating in the presence of the ionising background photo-evaporates the gas in DM halos, as described above. 
The bump in the PDF around $1 \lesssim \log_{10} \Delta \lesssim 2$ therefore disappears (red dot-dashed histogram). 
Note that the redistribution of the baryons due to photo-heating also slightly 
increases the minimum overdensity that is present in the simulation. In Appendix~\ref{Sec:Appendix:PDF} we compare 
the PDF obtained from our default simulation to the fit provided by \cite{Miralda:2000}, which is often employed to compute the 
clumping factor in (semi-)analytical studies of the epoch of reionisation.
\par
\subsection{The clumping factor}
\label{Sec:ClumpingFactor}
In this section we demonstrate how the clumping factor, 
$C \equiv \langle \rho_{\rm{b}}^2 \rangle_{\rm IGM} / \langle \rho_{\rm{b}}\rangle^2$, depends on the definition of
the IGM and compute it for our default simulation \textit{r9L6N256} and for the simulation \textit{L6N256}. This allows us to investigate  
how the clumping factor is affected by the inclusion of a photo-ionising background.
\par
Our main motivation for computing the clumping factor of the IGM is 
to evaluate the critical SFR density required to keep the IGM ionised. 
The critical SFR density describes the balance between the number of ionising photons 
escaping into the IGM (parametrised by the escape fraction) and the number 
of ionising photons that are removed from the IGM due to photo-ionisations of 
recombining hydrogen ions (parametrised by the clumping factor). When the ratio of photon escape rate
to recombination rate is larger than unity, the rate at which galaxies form stars exceeds 
the critical SFR density and is thus sufficient to keep the IGM ionised. 
\par
It is important to realise that 
only recombinations leading to the removal of ionising photons {\it which
  escaped the ISM of the star-forming regions} 
contribute to the balance that gives rise to the definition of the critical SFR density. 
To separate the gas in the ISM from the gas in the IGM, a simple threshold
density criterion is often employed (e.g.~\citealp{Miralda:2000}; see also the
discussion in \citealp{Miralda:2003}). Ionising
photons are counted as escaped once they enter regions with gas densities
$\rho_{\rm b} < \rho_{\rm thr}$. Consequently, only gas with densities $\rho_{\rm b} <
\rho_{\rm thr}$, or equivalently, gas with overdensities $\Delta < \Delta_{\rm thr} \equiv \rho_{\rm thr} / \langle \rho_{\rm b}\rangle$
should be considered in the evaluation of the clumping factor.
\par
The threshold density $\rho_{\rm thr}$ depends on which gas is considered to be
part of the ISM, and which gas is considered to be part of the IGM. As long as
the definition of the escape fraction and that of the clumping factor refer to
the same decomposition of the gas into IGM and ISM, its value can be chosen arbitrarily. We therefore 
treat the threshold density as a parameter and compute the clumping factor as a function
of $\Delta_{\rm thr}$ (cp.~\citealp{Miralda:2000}), 
\begin{equation}
  C(<\Delta_{\rm thr}) \equiv  \int_0^{\Delta_{\rm thr}} d\Delta\ \Delta^2~\mathcal{P}_{\rm V}(\Delta),
  \label{Eq:Clumping:Definition}
\end{equation}
where $\mathcal{P}_{\rm V}(\Delta)$ is normalised according to $\int_0^{\Delta_{\rm thr}} d\Delta\ \mathcal{P}_{\rm V}(\Delta) = 1$. 
In practice, we calculate $C(< \Delta_{\rm thr})$  by performing a volume-weighted 
summation over all SPH particles with overdensities 
$\Delta_i < \Delta_{\rm thr}$, i.e 
\begin{equation}
C(<\Delta_{\rm thr})= \frac{\sum_{\Delta_i < \Delta_{\rm thr}}  h_i^3 \Delta_i^2} { \sum_{\Delta_i < \Delta_{\rm thr}} h_i^3}, 
\end{equation}
where $h_i$ is the radius of the SPH smoothing kernel associated with SPH particle $i$. 
We verified that replacing $h_i^3$ with $m_i/\rho_i$ as an estimate for the volume occupied by SPH particle 
$i$ (with mass $m_i$) gives nearly indistinguishable results.
\par
By definition, $C(< \Delta_{\rm thr})$ increases monotonically with the threshold
density $\rho_{\rm thr}$. Here, we set an upper limit to $\Delta_{\rm thr}$, corresponding to the
threshold density $n^*_{\rm H} \equiv 10^{-1}\cmci$ for the onset of star formation that we employ in our
simulations. 
Since we impose an effective EoS for gas with densities larger than $n^*_{\rm H}$ (Section~\ref{Sec:Simulations}),
its PDF is not expected to reflect the PDF of real star-forming regions, 
motivating our choice for the maximum threshold density. The choice is conservative, leading 
to an overestimate rather than an underestimate of the critical SFR density, since the threshold density marking the escape of ionising photons and 
hence the clumping factor of the IGM to be used in Eq.~\ref{Eq:CriticalSfr} is likely to be lower (see, e.g., the discussion in \citealp{Gnedin:2008b}).    
\par
In the right-hand panel of Fig.~\ref{Fig:Cmax} we show $C(< \Delta_{\rm thr})$ for the simulations \textit{r9L6N256} and \textit{L6N256} 
at redshifts shortly before (at $z = 9.08$, when \textit{r9L6N256} 
and \textit{L6N256} are identical) and well after (at $z = 6$, when they differ by the presence and absence of an ionising 
background, respectively) the reheating redshift $z_{\rm{r}} = 9$. In agreement with our discussion above, the clumping 
factor increases monotonically with the threshold density. Its dependence on redshift can be understood by
looking at the evolution of the shape of the PDF, which we discussed in the previous section.
\par
For \textit{L6N256}, i.e. in the absence of photo-heating, the clumping factor for threshold overdensities $\log_{10} \Delta_{\rm thr} > 1$ 
is larger at $z = 6$ than at $z = 9.08$, due mainly to the growth of the bump present in the PDF for overdensities
$1 \lesssim \log_{10} \Delta \lesssim 2$. For $\log_{10} \Delta_{\rm thr} \sim 0$, 
on the other hand, the clumping factor is slightly smaller 
at $z = 6$ than at $z = 9.08$, which is caused by the depletion of the diffuse IGM through accretion onto DM halos.
Note that at $z = 6$ the clumping factor reaches a maximum value of $C \approx 20$, which is significantly smaller than the value quoted by
\cite{Gnedin:1997}, which is commonly employed in observational studies. 
This is probably because \cite{Gnedin:1997} computed the clumping factor including gas of any density, i.e. using a density
threshold implicitly set by the maximum overdensity resolved in their simulation. 
\par
The evolution of the clumping factor in \textit{r9L6N256}, i.e. in the presence of the ionising background, is very different.
At $z = 6$ it is close to that at  $z = 9.08$ for all threshold overdensities.
Compared to \textit{L6N256}, the difference between the clumping factors for $z = 6$ and $z = 9.08$ is greatly reduced and the clumping factor at redshift $z = 6$ never reaches values larger than $C \approx 6$.
\par
\subsubsection{Convergence tests}
\label{Sec:Convergence}
\begin{figure}
  \includegraphics[trim = 5mm 10mm 0mm 20mm, width=0.47\textwidth, clip=true]{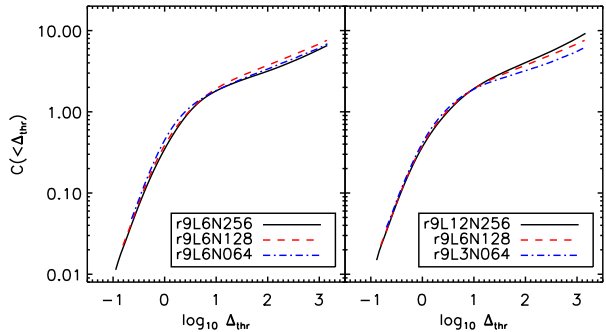} 
  \caption{Clumping factor $C(< \Delta_{\rm thr})$ of gas with overdensities
    $\Delta < \Delta_{\rm thr}$ 
    and its dependence on resolution (at fixed box size, left-hand panel) and on box size (at fixed 
    resolution, right-hand panel). The clumping factor obtained from our default simulation \textit{r9L6N256} 
    is converged with respect to the employed resolution for all threshold overdensities shown. With respect to
    the size of the simulation box, it is converged for threshold overdensities $\log_{10}\Delta_{\rm thr} \lesssim 2$.
    For larger threshold overdensities, full convergence may require the use of simulation boxes even larger than $12.5 \cMpch$,
    the size of the largest box employed here.}
  \label{Fig:Resolution}
\end{figure}
In this section we check whether our results are converged. Generally, one expects the clumping factor to increase with both the
spatial resolution and the size of the simulation box. The spatial resolution determines the smallest scale on which 
fluctuations in the density field may be identified, whereas the size of the simulation box sets a cut-off to the largest scale
on which the overdensity field can be non-zero. Moreover, the size of the simulation box limits the mass of the 
largest halo present in the simulation. 
Fig.~\ref{Fig:Resolution} demonstrates that our default simulation (\textit{r9L6N256}) is of sufficiently high resolution
and employs a sufficiently large box to allow a faithful computation of the clumping factor of the reheated IGM at $z = 6$.
In the left-hand panel we show the clumping factor in three simulations that
use the same box size, but have mass resolutions 
that differ by multiples of $2^3$, whereas in the right-hand panel we show the clumping factor in three simulations that employ the same
resolution, but have box sizes that differ by multiples of 2.
\begin{figure*}
\includegraphics[width=0.49\textwidth]{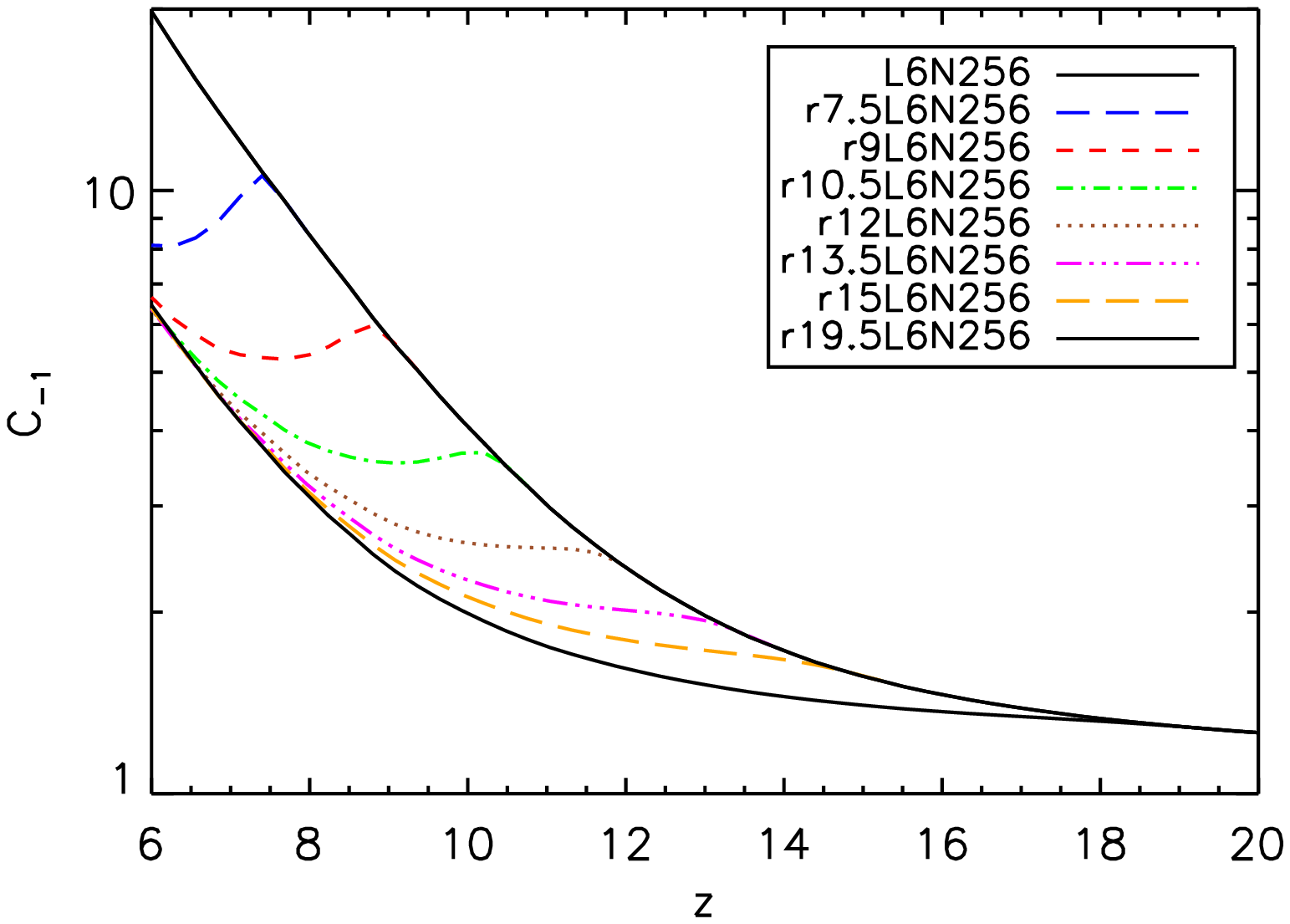}
  \includegraphics[width=0.49\textwidth]{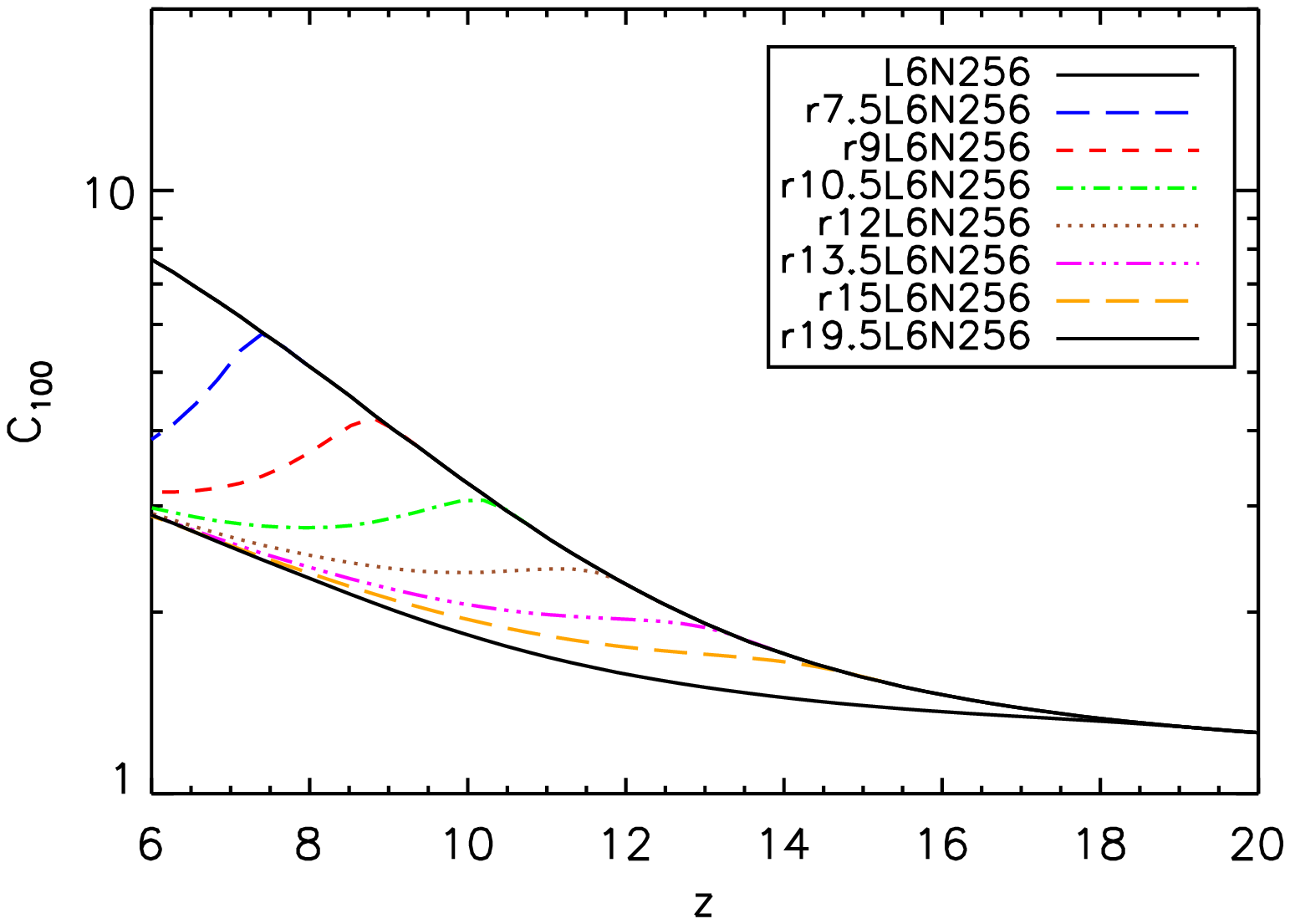}
  \caption{Evolution of the clumping factors $C_{-1}$ (left-hand panel) and $C_{100}$ (right-hand panel) for different 
    reheating redshifts $z_{\rm{r}}$, as indicated in the legends. Note that
    $C_{-1}(z)$ and $C_{100}(z)$ in all reheating simulations evolve towards
    the clumping factors obtained in the simulation with reheating at $z_{\rm r} = 19.5$ (bottom black solid curves).  
    At $z = 6$, the clumping factors are therefore 
    insensitive to the reheating redshift provided that $z_{\rm r} \gtrsim 9$, 
    with  $C_{-1}(z = 6) \approx 6$ and  $C_{100}(z = 6) \approx 3$. 
    Fits to the evolution of the clumping factors are given in Appendix~\ref{Sec:Appendix:Clumping}.
  }
\label{Fig:Clumping:Z}
\end{figure*}
\par
When compared to our default simulation \textit{r9L6N256}, 
decreasing the mass resolution by factors of $8$ (\textit{r9L6N128}) and $64$ (\textit{r9L6N064}) 
only leads to insignificant and unsystematic\footnote{Note that the clumping factor is even slightly larger in
\textit{r9L6N128} than in \textit{r9L6N256}, although the latter has an
$8$ times higher mass resolution.} changes in the clumping factor (left-hand panel).
This can be understood by noting that the virial mass of halos corresponding 
to a virial temperature $T_{\rm vir} = T_{\rm r}$ is 
resolved with $\gtrsim 100$ particles for redshifts $z < 9$. Any further increase 
in resolution would therefore mostly affect the abundance of DM halos with  $T_{\rm vir} \ll T_{\rm r}$. In the presence of the UV background
the gas in these halos is, however, quickly photo-evaporated (if formed at $z > z_{\rm r}$) or prevented from accreting by the 
large Jeans mass associated with the photo-heated gas (for $z < z_{\rm r}$). 
At redshifts well below the reheating redshift $z_{\rm r}$, i.e. when sufficient time has passed to accomplish the 
photo-evaporation of halos and to allow the gas to respond hydrodynamically to the jump in the Jeans mass
(i.e. to Jeans-filter the IGM), the observed convergence with respect to mass resolution 
is therefore expected\footnote{\label{footnote}We note that in the absence of a photo-ionising background, 
convergence may be more difficult to achieve, requiring a higher mass resolution than employed here. Convergence will be easier to achieve for lower 
threshold densities.}. 
\par
Since the clumping factor 
has already converged for the spatial resolution used in simulation \textit{r9L6N128}, we can employ this simulation
to verify whether the size of the box of our default simulation \textit{r9L6N256} is sufficiently large 
to enable an unbiased estimate of the clumping factor (right-hand panel). Overall, increasing the
box size (at fixed resolution) from $6.25 \cMpch$ by a factor of two to
$12.5 \cMpch$ leaves the clumping factor almost unaffected. For threshold densities $\log_{10}\Delta_{\rm thr} \gtrsim 2$ 
the clumping factor may, however, not yet have fully converged, indicating that even larger simulation boxes than that considered 
here may be required for its computation (see also the discussion in \citealp{Barkana:2004}).
\par
\subsubsection{Varying the reheating redshift}
\label{Sec:ReheatingRedshift}
To study the effect of photo-heating on the evolution of the clumping factor
in more detail, we make use of the clumping factors
\begin{equation}
  C_{-1} \equiv C( < 10^{-1} \cmci m_{\rm H} / (X \langle \rho_{\rm b}\rangle))
\end{equation}
and
\begin{equation}
  C_{100} \equiv C(< \min (100, 10^{-1} \cmci m_{\rm H} / (X \langle \rho_{\rm b} \rangle))). 
\label{Eq:C100}
\end{equation}
$C_{-1}$ is the clumping factor for gas with densities below $n_{\rm H} = n_{\rm H}^* \equiv 10^{-1}\cmci$,  the maximum threshold density
we consider. For redshifts $z < 16.3$,  $C_{100}$ fixes the threshold overdensity to a value between 
the mean overdensity of spherical top-hat DM halos, ($\approx 18 \pi^2$; e.g.~\citealp{Padmanabhan:1993}) 
and the overdensity at the virial radius of an isothermal DM halo
($\approx 60$; \citealp{Lacey:1994}) and closely agrees with the threshold densities commonly employed in the literature 
to calculate the clumping factor of the IGM. For larger redshifts, the threshold overdensity $\Delta_{\rm thr}= 100$ corresponds
to a density $n_{\rm H} > n_{\rm H}^* \equiv 10^{-1} \cmci$, which is larger than the critical density for the onset of star formation in our simulations. 
The definition Eq.~\ref{Eq:C100} ensures that the maximum density that we consider for the computation of $C_{100}$ is always smaller 
than $n_{\rm H}^*$. Note that $C_{-1}$ refers to the clumping factor of gas with densities below a fixed proper density, while
$C_{100}$ refers to the clumping factor of gas with densities below a fixed overdensity for redshifts $z < 16.3$ and is identical to
$C_{-1}$ for larger redshifts.
\par
The evolution of $C_{-1}$  and $C_{100}$ is shown, respectively, in the left-hand and right-hand panels of Fig.~\ref{Fig:Clumping:Z} for
the simulations \textit{r9L6N256} (i.e. reheating at redshift $z_{\rm r} = 9$)
and \textit{L6N256} (i.e. no reheating).
In the same figure we also include the evolution of  $C_{-1}$  and $C_{100}$ obtained
from the set of simulations \textit{r[$z_{\rm r}$]L6N256}, where $z_{\rm r} = 7.5, 10.5, 12, 13.5, 15$ and $19.5$.
While the simulations are identical for $z > z_{\rm{r}}$, 
the ionising background strongly affects the evolution of $C_{-1}(z)$  and
$C_{100}(z)$ for $z < z_{\rm{r}}$. 
Whereas in \textit{L6N256} the clumping factors reach $C_{-1} \approx 20$ and $C_{100} \approx 8$ at $z = 6$,
they only reach $C_{-1} \approx 6$ and $C_{100} \approx 3$ in \textit{r9L6N256}, which are smaller than in \textit{L6N256} 
by roughly a factor of three. Note that because the clumping factor obtained from simulation \textit{L6N256} is likely to be not fully converged
with respect to resolution (see footnote~\ref{footnote}), we may even have underestimated 
the magnitude of the decrease in the clumping factor due to photo-heating.
\par
Interestingly, the clumping factor at $z = 6$ 
is insensitive to the redshift $z_{\rm r}$ at which the UV background is turned on,
as long as $z_{\rm r} \gtrsim 9$. This is because, after an initial transitory phase,
the evolution of the clumping factor obtained for reheating at redshift $z_{\rm r}$ approaches that obtained 
for reheating at $z_{\rm r} = 19.5$. Note that the difference between the clumping factors 
obtained from \textit{r[$z_{\rm r}$]L6N256} and \textit{r19.5L6N256} becomes smaller with increasing reheating redshift $z_{\rm r}$. 
In particular, the clumping factors obtained for reheating at $z_{\rm r} = 15$ are nearly identical to those obtained 
for reheating at $z_{\rm r} = 19.5$ 
at all redshifts. The evolution of the clumping factors obtained from \textit{r19.5L6N256} 
can therefore be considered to reflect the evolution of the clumping factor in the limit of reheating at very high redshift, 
$z_{\rm r} \gg 19.5$. 
\par
In Appendix~\ref{Sec:Appendix:Clumping} we provide fits to the evolution of the clumping factor over the redshift range $6 \le z \le 20$ 
for a range of (over-)density thresholds and reheating redshifts.
These fits (Eqs.~\ref{Eq:Fit:C1} and \ref{Eq:Fit:C2}) may be employed in
(semi-)analytic models of the epoch of reionisation. Many such models
assume that reionisation heating provides only a negative feedback on the
reionisation process, reducing the star formation rate due to the
photo-evaporation of gas in low-mass halos.
However, as we have shown here, photo-heating decreases the clumping
factor, and hence the average recombination rate. Since this makes it easier
to keep the IGM ionised, reionisation heating also provides a positive feedback on the process of reionisation. 
Although the relative importance of both can only be assessed using larger
hydrodynamical simulations of higher resolution\footnote{At $z = 6$, 
  the cosmic SFR density (the stellar mass) in \textit{r9L6N256} is smaller than that in 
  \textit{L6N256} by a factor $1.26$ ($1.17$). In our simulations, photo-heating thus decreases the SFR density
  less strongly than it reduces the clumping factor, which would imply that the positive
  feedback is more important. The SFR densities in both \textit{r9L6N256} and \textit{L6N256} 
  are, however, not fully converged with respect to resolution and box size. A final assessment as to whether the
  negative or the positive feedback is stronger must therefore be deferred to future
  studies using simulations with higher resolution and larger box sizes.}, it is clear 
that models that do not account for this positive feedback will 
underestimate the efficiency with which star-forming galaxies are able to reionise the IGM.
\par
\subsubsection{Dependence on the reheating temperature}
\label{Sec:ReheatingTemperature}
\begin{figure}
  \includegraphics[width=0.47\textwidth, clip=true]{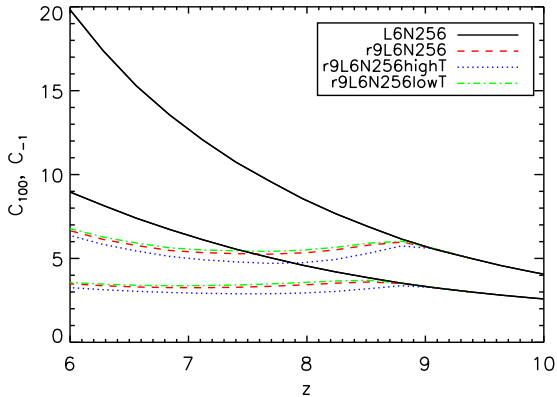} 
  \caption{The dependence of $C_{-1}$ (upper set of curves) and $C_{100}$ (lower set of curves) 
    on the value for the reheating energy $\epsilon_{\rm r}$. Both \textit{r9L6N256highT} ($\epsilon_{\rm r} = 20 \eV$) 
    and \textit{r9L6N256lowT} ($\epsilon_{\rm r} = 0 \eV$) give results very similar to that obtained from 
  the default run,  \textit{r9L6N256} ($\epsilon_{\rm r} = 2 \eV$), which demonstrates
  the robustness of our conclusions with respect to changes in the reheating temperature.}
  \label{Fig:CTemp}
\end{figure}
In this section we investigate the robustness of our results with respect to our simplified treatment of photo-ionisation 
heating. 
\par
We have considered photo-heating by a uniform ionising background in the optically thin limit.  
In reality, the reionisation process is likely to be driven by inhomogeneously distributed sources in an initially
optically thick medium. The temperature to which the IGM is reheated will then not only depend on the spectrum of the ionising sources,
but also on the amount of spectral hardening due to the preferential
absorption of the less energetic ionising photons 
(e.g.~\citealp{Abel:1999}; \citealp{Bolton:2004}). Moreover, the speed at which a particular patch of the IGM
is reionised determines the duration during which its gas can cool efficiently, as the cooling is dominated by inelastic
collisions between free electrons and neutral atoms. Different reionisation histories may therefore 
result in different IGM temperatures (e.g.~\citealp{Miralda:1994}; \citealp{Theuns:2002}; \citealp{Hui:2003}; \citealp{Tittley:2007}). 
\par
To bracket possible scenarios, we have performed two additional simulations 
in which we varied the amount of energy transferred to the baryons during the  photo-ionisation process,
\textit{r9L6N256highT} and \textit{r9L6N256lowT}. Whereas in the former we employ an additional energy 
input that is ten times larger than our default value ($\epsilon_{\rm r} = 20 \eV$), 
in the latter no additional energy is injected ($\epsilon_{\rm r} = 0 \eV$). We show in Fig.~\ref{Fig:CTemp}
that the evolution of $C_{-1}$ and $C_{100}$ obtained from these two simulations is very similar to that obtained 
from our default run, \textit{r9L6N256}. The dependence on the reheating
temperature $T_{\rm r} > 10^4 \K$ is weak, because halos with virial
temperatures $T_{\rm vir} \lesssim 10^4 \K$ are already efficiently destroyed for 
$T_{\rm r} \approx 10^4 \K$. A further increase in the reheating temperature
mostly affects the fraction of mass in halos with larger virial temperatures, which is small. Moreover, Fig.~\ref{Fig:ThermEvol} shows
that the gas in the simulations \textit{r9L6N256highT} and \textit{r9L6N256lowT}
quickly loses memory of its thermal state at some higher redshift, which is another 
reason for the similarity in the results obtained using different values for the reheating energy $\epsilon_{\rm r}$.
\subsubsection{Effect of kinetic supernova feedback and dependence on cosmological parameters}
\label{Sec:Feedback}
The inclusion of kinetic feedback from supernovae in \textit{r9L6N256winds}
only weakly affects the evolution of the clumping factors. At redshift $z = 6$, $C_{-1}$ and $C_{100}$ 
are slightly larger (by factors $1.1$ and $1.18$, resp.)  in \textit{r9L6N256winds} 
than in our default simulation, \textit{r9L6N256}, which does not include kinetic feedback. The
reason for the slight increase in the clumping factors is that winds move gas 
from regions of densities larger than the critical density for the
onset of star formation to regions of lower density that 
contribute to the calculation of the clumping factors. We note that the inclusion of kinetic feedback does, on the other hand, 
strongly affect the cosmic SFR. At $z = 6$, the cosmic SFR (the stellar mass) is lower in the simulation that includes kinetic feedback 
(\textit{r9L6N256winds}) than in our default simulation (\textit{r9L6N256}) by a factor of $6.1$ ($3.4$).
\par
Finally, we quote the clumping factors obtained from the simulations $r9L6N256W3$ and  $r9L6N256W1$,
which employed cosmological parameters consistent with the WMAP 3-year and 1-year observations, respectively.
We find that at redshift $z = 6$, the clumping factors $C_{-1}$ and $C_{100}$ are larger in  $r9L6N256$ than in 
$r9L6N256W3$ by factors of $1.31$ and $1.16$, respectively. They are smaller in $r9L6N256$ than in 
$r9L6N256W1$ by factors of $0.74$ and $0.84$. In summary, with respect to $r9L6N256$, the clumping factors are larger in $r9L6N256W1$
and smaller in $r9L6N256W3$, as expected from the corresponding values of $\sigma_8$, which set the 
average absolute amplitude of the overdensity fluctuations.
\par
\subsubsection{Comparison with previous work}
\begin{figure}
  \includegraphics[width=0.47\textwidth, clip=true]{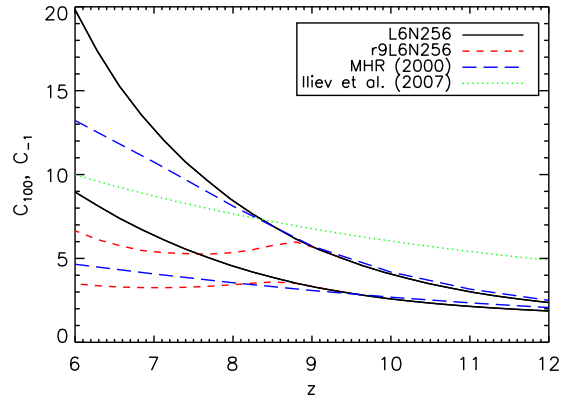} 
  \caption{Clumping factor evolution: comparison with previous work. The black
  solid and red dashed curves are the clumping factors $C_{-1}$ (upper set of
  curves) and $C_{100}$ (lower set of curves) obtained from our simulations
  \textit{L6N256} and \textit{r9L6N256}. The 
  blue dashed curves show the evolution of the clumping factors $C_{-1}$
  (upper curve) and  $C_{100}$ (lower curve) derived from the gas density PDFs
  presented in \protect\cite{Miralda:2000}, which implicitly incorporate the effects of
  photo-heating. The green dotted curve shows the evolution of the clumping
  factor (defined without explicitly referring to a (over-)density threshold;
  instead, the overdensity threshold was set by the numerical resolution) the dark matter simulation of \protect\cite{Iliev:2007}, which
  does not include the effects of photo-heating. In both cases a direct
  comparison is difficult, because of the different assumptions underlying the
  individual works.}
  \label{Fig:Comparison}
\end{figure}

We conclude our study of the clumping factor with a brief comparison
with previous work, shown in Fig.~\ref{Fig:Comparison}. The evolution of the
clumping factors in our simulations \textit{L6N256} and \textit{r9L6N256} is shown by the black solid 
and red dashed curves, respectively, where the upper (lower) set of curves shows $C_{-1}$ ($C_{100}$).
We compare it to the evolution of the clumping factor presented in 
\cite{Miralda:2000} and \cite{Iliev:2007}, which are amongst the most commonly employed
works on the clumping factor and make use of sufficiently different techniques to
bracket a range of possible cases. We caution the reader that such a direct
comparison is difficult and of limited validity because of the 
very different assumptions underlying the individual works.
\par
\cite{Miralda:2000} used the L10 hydrodynamical simulation presented in
\cite{Miralda:1996} to obtain the PDF of the gas
density at redshifts $z = 2, 3$ and $4$. The simulation was performed using
the TVD hydrodynamical scheme described in \cite{Ryu:1993}. It used a box of size
$10\cMpch$, $144^3$ dark matter particles and $288^3$ gas cells and employed
cosmological parameters consistent with the first-year COBE normalization. The
simulation included photo-heating from a uniform UV
background, computed from the emissivities of the sources in the simulation. 
We refer the reader to \cite{Miralda:1996} for more details. 
\cite{Miralda:2000} also provided fits to the gas density PDF and presented a
prescription for its extrapolation to redshifts $z > 4$. We employed this
prescription to compute the clumping factor evolution using Eq.~\ref{Eq:Clumping:Definition}. 
\par
The evolution of the clumping factors $C_{-1}$ and $C_{100}$ obtained from the \cite{Miralda:2000} 
PDFs is shown, respectively, by the top and bottom blue long-dashed curves. For redshifts $z \gtrsim 9$, it
closely agrees with the corresponding evolution obtained from our
simulation \textit{r9L6N256}. For lower redshifts the agreement is less good, although the clumping
factors never differ by more than factors $\sim 2$. The
differences between their and our results are probably due to the use of
different hydrodynamical schemes, different cosmological parameters and
different prescriptions for the UV background. The change in the slope of the
clumping factor growth that can be seen at redhift $z \approx 9$ is likely due to the inclusion of photo-heating. 
That this change is much less pronounced than in our simulation \textit{r9L6N256} may be due to
a more gradual build-up of the ionising background in the \cite{Miralda:2000} simulation.
\par
\cite{Iliev:2007} computed the clumping factor from a pure dark matter 
simulation. The simulation employed a box of size $3.5 \cMpch$ and $1624^3$ particles and was
initialized with cosmological parameters consistent with the WMAP 3-year
results\footnote{We note that they also computed the clumping factor in a
  similar simulation that was initialized with cosmological parameters consistent
  with the WMAP 1-year results.}. The clumping
factor was computed by averaging over all dark matter densities, and hence
only implicitly makes use of an overdensity threshold (determined by the
maximum overdensity present in their simulation). 
We refer the reader to the original description in \cite{Iliev:2007} for more details.
\par
\cite{Iliev:2007} provided the following fit to the evolution of the clumping
factor in their simulation,
\begin{equation}
C_{\rm Iliev07}(z) = 26.2917 \exp (-0.1822 z+0.003505z^2),
\label{Eq:Clumping:Iliev}
\end{equation}
which is valid over the range $6 < z < 30$. It is shown by the green dotted curve.
Since it was derived from a pure dark matter simulation, Eq.~\ref{Eq:Clumping:Iliev} does not capture the
hydrodynamical response due to reionisation heating. It should therefore
be compared to the evolution of the clumping factor obtained from our simulation \textit{L6N256}, which did not
include photo-heating. A direct interpretation of such a comparison is, however, difficult, 
because Eq.~\ref{Eq:Clumping:Iliev} does not explicitly refer to an
overdensity threshold.
\par
Our comparison clearly illustrates that there is a considerable spread in the clumping
factor values quoted in the literature. The interpretation of many studies is complicated by the fact that they do not
refer to a density threshold, which means that the result is determined by the numerical resolution of their simulations.
\par
\section{Discussion}
\label{Sec:Conclusions}
Several observational studies have claimed that the star formation rate (SFR) density at redshift $z \approx 6$ 
is smaller than the critical SFR density required to keep the intergalactic 
medium (IGM) ionised. In the absence of a large population of unseen sources of ionising radiation, 
this discrepancy between the two SFR densities would be in direct conflict with the high degree of ionisation inferred from 
the non-detection of a Gunn-Peterson trough in the majority of the line-of-sight spectra towards $z \lesssim 6$ quasars.
\par
The critical SFR density is inversely proportional to the spatially averaged fraction of ionising photons that escape 
into the IGM per unit time and proportional to the clumping factor $C \equiv \langle \rho_{\rm{b}}^2 \rangle_{\rm IGM} /
\langle \rho_{\rm{b}}\rangle^2$, a measure for the average recombination rate in the IGM. 
One may therefore ask whether the discrepancy between the observed and critical SFR densities could be resolved
by changing the assumptions about the values of either of these two quantities.
In this work we considered the hypothesis that most observational studies 
overestimate the critical SFR density because they employ a clumping factor that is too large.
\par
We re-evaluated the clumping factor, analysing the gas density distributions in a set of cosmological smoothed 
particle hydrodynamics simulations that include radiative cooling and photo-ionisation by a uniform UV background
in the optically thin limit. The clumping factor of the IGM depends critically on 
the definition of which gas is considered to be part of the IGM. Following \cite{Miralda:2000}, we assumed that all gas 
with densities below a threshold density constitutes the IGM and computed the clumping factor as a function of 
this threshold density. In addition, we introduced two physically well-motivated definitions, $C_{100}$, 
the clumping factor of gas with overdensities $\Delta < 100$ and $C_{-1}$, the clumping factor
of gas with proper densities below  $n_{\rm H} = n_{\rm{H}}^* \equiv 10^{-1}\cmci $, our threshold density for the onset of star formation.
\par
By comparing simulations that include photo-ionisation by a uniform UV background  
to one that does not, we showed that photo-heating strongly influences the
evolution of the clumping factor of the IGM.
Photo-ionisation heating expels the gas from within halos of virial temperatures $T_{\rm vir} \lesssim 10^4 \K$ 
and prevents its further accretion by raising the Jeans mass in the IGM.
By suppressing the formation of stars in these low-mass halos, photo-heating from reionisation decreases
the rate at which ionising photons are emitted into the IGM and is therefore 
correctly said to exert a negative feedback on the reionisation process. 
The fact that photo-heating also leads to a decrease in the clumping factor 
and hence provides a strong positive feedback by making it easier to keep the IGM ionised, 
is however often overlooked (but see, e.g., \citealp{Haiman:2001}; \citealp{Oh:2003}; \citealp{Kuhlen:2005}; 
\citealp{Wise:2005}; \citealp{Furlanetto:2006}; \citealp{Ciardi:2007}).
\par
At redshift $z = 6$, we find that $C_{-1} \approx 6$ and $C_{100} \approx 3$ and that
these values are insensitive to the redshift $z_{\rm r}$ at which the UV background is turned on, as long as $z_{\rm r} \gtrsim 9$. These values for 
$C_{-1}$ and $C_{100}$ are at least three times smaller than they would be in the
absence of photo-heating.
We demonstrated that our default simulation is converged at $z = 6$ with respect to the employed resolution. 
It is converged with respect to changes in the box size for threshold overdensities $\log_{10}\Delta_{\rm thr} \lesssim 2$.
In Appendix~\ref{Sec:Appendix} we provide fits to the evolution of the clumping factor 
for various (over-)density thresholds and reheating redshifts.
There we also compare the probability density function (PDF) of the gas densities at $z = 6$ obtained from our default simulation to the widely used 
fit provided by \cite{Miralda:2000}. We update their fitting parameters to best fit the PDF from our default simulation.
Finally, we compared our results for the clumping factor of the IGM to those obtained in previous works.
\par
Since even our most conservative estimate for the clumping factor ($C_{-1} \approx 6$) is 
five times smaller than the clumping factor that is usually employed to determine the capacity of star-forming
galaxies to keep the $z = 6$ IGM ionised, our results may have important implications for the 
understanding of the reionisation process. Setting $C = 6$ in Eq.~\ref{Eq:CriticalSfr}, the critical SFR density becomes
$\dot{\rho}_* = 0.005~f_{\rm esc}^{-1}\Msun \invyr \invMpc$. This is smaller than 
recent observational estimates for the SFR density at $z \approx 6$,
$\dot{\rho}_* = 0.022 \pm 0.004\Msun \invyr \invMpc$ (integrated to the observed $z \approx 6$ faint-end limit $L > 0.04 ~L_{z=3}^\ast$ and
dust-corrected; \citealp{Bouwens:2007}), provided that $f_{\rm esc} \gtrsim 0.2$. 
\par
Our study thus suggests that the observed population of star-forming galaxies
may be capable of keeping the IGM ionised, 
relaxing the tension between observationally inferred and critical SFR density in view of the 
observation of a highly ionised IGM at redshifts  $z \lesssim 6$. 
We note that at $z \approx 7$, the SFR density is estimated to be
$\dot{\rho}_* = 0.004 \pm 0.002 \Msun \invyr \invMpc$ (integrated to the
observed $z \approx 7$ faint-end limit $L > 0.2 ~L_{z=3}^\ast$ and
dust-corrected; \citealp{Bouwens:2008}), whereas the critical SFR density is
$\dot{\rho}_* = 0.008~f_{\rm esc}^{-1} \Msun \invyr \invMpc$ (using $C =
6$). The observed population of star-forming galaxies at $z \approx 7$ is
therefore not able to keep the IGM ionised. If the Universe were ionised by
this redshift, then the sources that were responsible remain to be discovered.
\par
We caution the reader that the comparison of the critical and observed SFRs is
subject to considerable uncertainty. First, the SFR inferred from UV galaxy
counts probably underestimates the true SFR, because these counts miss UV
galaxies fainter than the faint-end limit implied by their sensitivities. These 
galaxies may, however, significantly contribute to the UV luminosity
density if the faint-end slope of the UV luminosity function is sufficiently
steep. Complementary estimates of the
high-redshift star formation rate based on measurements of the high-redshift ($z = 4-7$) gamma ray burst
rate (\citealp{Yuksel:2008}) and measurements of the Lyman-alpha forest opacity at redshifts $z \sim 3$
(\citealp{Faucher:2008}) indeed suggest $z\sim 6$ SFRs that exceed those inferred
from UV galaxy counts by factors of a few.
\par
Second, the expression for the critical SFR (Eq.~\ref{Eq:CriticalSfr})
is only approximate. As already mentioned in the introduction, this expression
is based on equating the rate at which ionising photons
escape into the intergalactic medium to the rate at which the intergalactic
gas is recombining, both averaged in space. It neglects effects like the cosmological redshifting of photons below the ionisation
threshold energy and evolution of the ionising sources during a recombination time. Because the
cross-section $\sigma_{\rm HI} \sim \nu^{-3}$ for absorption of ionising photons
by neutral hydrogen decreases with increasing photon frequency $\nu$, these effects may become important 
for photons whose mean free path is comparable to the cosmic horizon. Our limit on the escape fraction 
required to keep the Universe at $z \approx 6$ ionised may
therefore only be accurate within a factor of a few.
\par
Our simulations demonstrate that radiation-hydrodynamical feedback due
to photo-ionisation heating plays a key role in shaping the properties
of the IGM at redshifts $z \gtrsim 6$.  We have studied
the impact of photo-ionisation heating on the clumping factor of the
IGM assuming a uniform ionising UV background in the optically thin
limit. In reality the reionisation process will, however, be more
complex. We have demonstrated the robustness of our conclusions with
respect to uncertainties in the temperature of the IGM resulting from
our simplified treatment of the reionisation heating, but there are
other factors whose importance is more difficult to assess.
\par
Our use of the optically thin approximation neglects self-shielding, a
radiative transfer effect due to which halos that would otherwise be
completely photo-evaporated could keep some of their gas (e.g.,
\citealp{Kitayama:2000}; \citealp{Susa:2004}; \citealp{Dijkstra:2004}). Since the
self-shielded gas remains neutral, it should be excluded when
computing the clumping factor. Self-shielding becomes important for
$N_{\rm HI} \gtrsim 10^{18} \cmsi$, which for self-gravitating gas
clouds corresponds to densities (\citealp{Schaye:2001}) $n_{\rm H}
\gtrsim 10^{-2} \cmci (\Gamma / 10^{-12} \invs)^{-1}$, where $\Gamma$
is the HI photo-ionisation rate.  In our discussion of the critical
SFR we have conservatively adopted the clumping factor $C_{-1}$,
defined using a threshold density $n_{\rm H} = 10^{-1} \cmci$, which
is an order of magnitude larger than the critical density for
self-shielding.
\par
Self-shielding might affect our results because it may lower the speed with which halos are photo-evaporated.
The work by \cite{Iliev:2005b} (in combination with the work by \citealp{Shapiro:2004}) 
shows that photo-evaporation times obtained in simulations that employ an optically thin UV background may differ
by factors of a few from those obtained in detailed radiation-hydrodynamical simulations. 
However, if any halos would resist photo-evaporation much longer than predicted by our 
approximate treatment of photo-heating, then the clumping factor of the IGM 
would be even lower, because self-shielding locks the gas that would otherwise 
contribute to the clumping factor of the IGM in its neutral state.
\par
If absorption by optically thick (self-shielded) clouds becomes important, 
then the mean free path of ionising photons may be set by the mean distance 
between these clouds (\citealp{Zuo:1993}) instead of by the opacity of the diffuse IGM.
In this case it might be appropriate to supplement the \citealp{Miralda:2000} model for the computation of the  
average recombination rate with a more direct account for these clouds as discrete photon sinks (e.g., \citealp{Iliev:2005a}; \citealp{Ciardi:2006}). 
The effect of absorption by optically thick clouds is, however, largely degenerate with the ionising efficiency of 
the population of star-forming galaxies, as explained in \cite{Iliev:2007}. If, on the one hand, these clouds are 
ionising sources themselves, then their contribution to the average recombination rate can be described by 
their escape fractions. If, on the other hand, these clouds do not host ionising 
sources, then their contribution to the average recombination can be accommodated by changes in 
the ionising efficiency of star-forming galaxies because of their biased clustering around these galaxies. The effects of 
optically thick clouds can therefore approximately be accounted for by adjusting the properties of the ionising sources.
\par
Self-shielding is only one example of the physical effects that we
are ignoring. The inclusion of metals and molecules, for instance, 
would increase the ability of the gas to cool, 
which may lead to an increase in the clumping factor (e.g.~\citealp{Maio:2007}). 
In the presence of photo-heating and for the threshold densities we employ 
to compute the clumping factors we however expect this effect to be very small. 
A more efficient cooling due to metals and molecules may also enable star formation and 
associated kinetic feedback in low-mass haloes with virial temperatures $T_{\rm vir} < 10^4 \K$. Both are processes
that we have ignored but which are likely to affect the clumping factor evolution (e.g.~\citealp{Wise:2008}).  
\par
The evolution of the clumping factor will also depend on the morphology
of the reionisation transition. Our treatment implicitly assumed that all gas with overdensities smaller than a threshold overdensity
is uniformly ionised, while all gas with larger overdensities is fully neutral. This picture probably only applies to the late stages
of reionisation, when individual ionised regions start to overlap and the only neutral gas that remains to be ionised is locked up in 
regions of high gas overdensities. Before overlap, other reionisation models may be more useful for the description of the clumping factor 
evolution. For example, \cite{Furlanetto:2005} point 
out that if the large-scale dense regions are ionised first, the clumping factor 
may be somewhat larger than one would otherwise expect, because the photons are initially confined to
these dense regions. Moreover, \cite{Furlanetto:2008} show
that in fossil ionised regions, that is, regions in which the gas
freely recombines, the clumping factor will generally be smaller than
for regions in photo-ionisation equilibrium, because the densest gas
which contributes most to the clumping factor becomes neutral first.
\par
The clumping factor is an important ingredient of (semi-)analytic
treatments of reionisation. It would therefore be highly desirable to
evaluate the approximations we have employed in our simplified treatment
of the photo-heating process using large high-resolution
radiation-hydrodynamical simulations of reionisation that include
cooling by metals and molecules and feedback from star formation. At
the moment such simulations are, however, not yet feasible.
\par
In fact, current state-of-the-art radiative transfer simulations
typically make use of clumping factors in their ``sub-grid'' modules
because they lack the resolution to resolve the clumpiness of the gas
directly. In addition, they typically do not include the effect of photo-heating.
In fact, many radiative transfer simulations ignore hydrodynamics altogether and assume the gas to trace the dark matter.
\par
With this work we
hope to have presented a conservative assessment of the clumping factor of the post-reionisation IGM that may provide a useful input to
future (semi-)analytic models and simulations of the reionisation process.

\section*{Acknowledgments} 
We thank Rychard Bouwens, Steven Furlanetto and Jordi
Miralda-Escud{\'e} for valuable communications. We thank the anonymous referee for his/her 
constructive contribution that greatly improved the discussion of our results.
We are grateful to Claudio Dalla Vecchia for help with running the simulations, Rob
Wiersma for useful discussions on the cooling tables and Marcel Haas
and Freeke van de Voort for a thorough reading of the draft.  The
simulations presented here were run on the Cosmology Machine at the
Institute for Computational Cosmology in Durham as part of the Virgo
Consortium research programme and on Stella, the LOFAR BlueGene/L
system in Groningen.  This work was supported by Marie Curie
Excellence Grant MEXT-CT-2004-014112.

\appendix 
\section{Fitting formulas} 
Here we provide fits to the evolution of the clumping factor for a range of overdensity 
thresholds and all reheating redshifts considered for use with (semi-)analytical models of reionisation. 
We also compare the 
probability density function of the gas densities at redshift $z = 6$ obtained from our default simulation to the fit provided by \cite{Miralda:2000}.
\label{Sec:Appendix}
\subsection{Clumping factor}
\label{Sec:Appendix:Clumping}
In this section we fit the evolution of the clumping factors $C_{-1}$  and $C_{100}$  over the redshift range 
$6 \le z \le 20$, based on the data presented in Fig.~\ref{Fig:Clumping:Z}. In addition, we give fits to the evolution of 
the clumping factors $C_{-2}$ and $C_{1000}, C_{500}, C_{200}$ and $C_{50}$, where $C_{-2} \equiv C( < 10^{-2} \cmci  m_{\rm H} / (X \langle \rho_{\rm b} \rangle))$ and 
$C_{1000} \equiv C(< \min (1000, 10^{-1} \cmci m_{\rm H} / (X \langle \rho_{\rm b} \rangle)))$ and similar for $C_{500}, C_{200}$ and $C_{50}$.
We first give fits to the evolution of the clumping factors for the simulations  \textit{L6N256} and
\textit{r19.5L6N256}, i.e. the simulations without reheating and with reheating at the highest redshift we considered. These fits are then used 
to obtain fits to the evolution of the clumping factors for reheating at the intermediate redshifts 
$z_{\rm r} = 7.5, 9.0, 10.5, 12, 13.5$ and $15.0$ by interpolation.
\par
We approximate the evolution of the clumping factors for the simulations \textit{L6N256} and 
\textit{r19.5L6N256} by  
\begin{eqnarray}
C(z) &=&  z^{\beta} e^{-\gamma z + \delta} + \alpha \label{Eq:Fit:C1},
\end{eqnarray}
where $C$ is either $C_{-1}, C_{-2}, C_{1000}, C_{500}, C_{200}, C_{100}$ or $C_{50}$ and similar for $\alpha$, $\beta$, $\gamma$, $\delta$.
The values for the parameters $\alpha$, $\beta$, $\gamma$, $\delta$
are listed in Tables~\ref{tbl:constants1} and \ref{tbl:constants2}. The fit (Eq.~\ref{Eq:Fit:C1}) is accurate to within $\lesssim 10 \%$.
We emphasize that it is only strictly valid over the fitting range $6 \le z \le 20$.
For $C_{1000}, C_{500}, C_{200}, C_{100}$ and  $C_{50}$, i.e. for the clumping factors that are defined using an overdensity threshold, we forced, however, 
the fits to approach the correct high-$z$ limit, i.e. $C \to 1$, by fixing $\alpha = 1$ during the fitting procedure. 
Note that the threshold densities used with $C_{-1}$ and $C_{-2}$ correspond to threshold overdensities $\Delta_{\rm thr} < 1$ for redshifts $z > 79.4$ and 
$z > 36.3$, respectively and that $C_{1000}, C_{500}, C_{200}, C_{100}$ and $C_{50}$ become identical to $C_{-1}$ for redshifts $z > 7.0, 9.1, 12.7, 16.3$ 
and $20.8$, respectively.
\par
The values of $C$ for reheating at redshifts $z_{\rm r} = 7.5, 9.0, 10.5, 12, 13.5$ and $15.0$
(hereafter $C^{\rm z_{\rm r}}$) are fitted by interpolating between 
the fits (Eq.~\ref{Eq:Fit:C1}) to the evolution of the clumping factors obtained from \textit{L6N256} (hereafter $C^0$) and 
\textit{r19.5L6N256} (hereafter $C^{19.5}$). That is, we write
\begin{equation}
  C^{z_{\rm r}}(z) = w(z) C^0(z) + [1 - w(z)] C^{19.5}(z) \label{Eq:Fit:C2},
\end{equation}
where 
\begin{equation}
  w(z) = \frac{1}{2} \left [\textrm{erf} \left (\frac{z - \zeta^{z_{\rm r}}}{\tau^{z_{\rm r}}}\right) + 1 \right],
\end{equation}
and $\textrm{erf}$ is the error function,
\begin{equation}
\textrm{erf} (z) = \frac{2}{\sqrt\pi} \int_0^z  d\tilde{z}\  \exp \left(-\tilde{z}^2\right).
\end{equation}
The constants $\zeta^{z_{\rm r}}$ and $\tau^{z_{\rm r}}$ are listed
in Tables~\ref{tbl:constants3} and \ref{tbl:constants4}. Eq.~\ref{Eq:Fit:C2} fits the data to within $\lesssim 10 \%$.
\par
\begin{table}
\begin{center}
  \caption{Parameter values $\alpha$, $\beta$, $\gamma$ and $\delta$ to be used in Eq.~\ref{Eq:Fit:C1} in order to fit the evolution of the
  clumping factors $C_{-1}$ and $C_{-2}$ obtained from the simulations \textit{L6N256} and \textit{r19.5L6N256} (see also Fig.~\ref{Fig:Clumping:Z}).
\label{tbl:constants1}}
\begin{tabular}{cccccccc}
\hline
\hline
 & \textit{L6N256} & \textit{r19.5L6N256} &  \\
\hline
$\alpha_{-1}$         & $1.29$ & $1.21$ &  \\
$\beta_{-1}$         & $0.00$ & $-3.66$ & \\
$\gamma_{-1}$         & $0.47$ & $0.00$ &  \\
$\delta_{-1}$         & $5.76$ & $8.25$ &  \\
\hline
$\alpha_{-2}$         & $1.29$ & $1.16$ &  \\
$\beta_{-2}$        & $0.00$ & $-2.47$ &  \\
$\gamma_{-2}$       & $0.44$ & $0.00$ &  \\
$\delta_{-2}$       & $4.68$ & $5.16$ &  \\
\hline
\end{tabular}
\end{center}
\end{table}
\begin{table}
\begin{center}
  \caption{Parameter values $\alpha$, $\beta$, $\gamma$ and $\delta$  to be used in Eq.~\ref{Eq:Fit:C1} in order to fit the evolution of the
    clumping factors $C_{1000}, C_{500}, C_{200}, C_{100}$ and $C_{50}$ obtained from the simulations \textit{L6N256} and \textit{r19.5L6N256} 
    (see also Fig.~\ref{Fig:Clumping:Z}).
    \label{tbl:constants2}}
\begin{tabular}{cccccccc}
\hline
\hline
 & \textit{L6N256} & \textit{r19.5L6N256} &  \\
\hline
$\alpha_{1000}$         & $1.00$ & $1.00$ &  \\
$\beta_{1000}$         & $-1.00$ & $-2.89$ & \\
$\gamma_{1000}$         & $0.28$ & $0.00$ &  \\
$\delta_{1000}$         & $6.29$ & $6.76$ &  \\
\hline
$\alpha_{500}$         & $1.00$ & $1.00$ &  \\
$\beta_{500}$        & $0.00$ & $-2.44$ &  \\
$\gamma_{500}$       & $0.34$ & $0.00$ &  \\
$\delta_{500}$       & $4.60$ & $5.68$ &  \\
\hline
$\alpha_{200}$         & $1.00$ & $1.00$ &  \\
$\beta_{200}$         & $0.00$ & $-1.99$ & \\
$\gamma_{200}$         & $0.30$ & $0.00$ &  \\
$\delta_{200}$         & $4.04$ & $4.49$ &  \\
\hline
$\alpha_{100}$         & $1.00$  & $1.00$ &  \\
$\beta_{100}$        & $0.00$ & $-1.71$ &  \\
$\gamma_{100}$       & $0.28$ & $0.00$ &  \\
$\delta_{100}$       & $3.59$ & $3.76$ &  \\
\hline
$\alpha_{50}$         & $1.00$ & $1.00$ &  \\
$\beta_{50}$        & $0.00$ & $-1.47$ &  \\
$\gamma_{50}$       & $0.23$ & $0.00$ &  \\
$\delta_{50}$       & $2.92$ & $3.08$ &  \\
\hline
\end{tabular}
\end{center}
\end{table}

\begin{table}
\begin{center}
\caption{Parameter values $\zeta^{z_{\rm r}}$ and $\tau^{z_{\rm r}}$ 
  to be used in Eq.~\ref{Eq:Fit:C2} in order to fit the evolution of the clumping factors  $C_{-1}$ and $C_{-2}$
  obtained from the simulations \textit{r[$z_{\rm{r}}$]L6N256} (see also Fig.~\ref{Fig:Clumping:Z}).
\label{tbl:constants3}}
\begin{tabular}{cccccccc}
\hline
\hline
$ z_{\rm r}$ & $\zeta^{z_{\rm r}}_{-1}$ & $\tau^{z_{\rm r}}_{-1}$  & $\zeta^{z_{\rm r}}_{-2}$ & $\tau^{z_{\rm r}}_{-2}$ &  \\
\hline
7.5        & $6.83$ & $0.83$ & $6.61$ & $0.80$ &  \\
9.0        & $8.10$ & $1.25$ & $7.78$ & $1.26$ &  \\
10.5        & $9.41$ & $1.71$ & $8.92$ & $1.65$ &  \\
12.0        & $10.77$ & $2.16$ & $10.02$ & $1.96$ &  \\
13.5        & $12.10$ & $2.45$ & $11.09$ & $2.24$ &  \\
15.0        & $13.27$ & $2.43$ & $12.20$ & $2.48$ &  \\
\hline
\end{tabular}
\end{center}
\end{table}

\begin{table*}
\begin{center}
\caption{Parameter values $\zeta^{z_{\rm r}}$ and  $\tau^{z_{\rm r}}$ 
  to be used in Eq.~\ref{Eq:Fit:C2} in order to fit the evolution of the clumping factors  $C_{1000}, C_{500}, C_{200}, C_{100} $ and $C_{50}$
  obtained from the simulations \textit{r[$z_{\rm{r}}$]L6N256} (see also Fig.~\ref{Fig:Clumping:Z}).
\label{tbl:constants4}}
\begin{tabular}{ccccccccccccc}
\hline
\hline
$ z_{\rm r}$ & $\zeta^{z_{\rm r}}_{1000}$  & $\tau^{z_{\rm r}}_{1000}$ &  $\zeta^{z_{\rm r}}_{500}$  & $\tau^{z_{\rm r}}_{500}$ & $\zeta^{z_{\rm r}}_{200}$  & $\tau^{z_{\rm r}}_{200}$ &  $\zeta^{z_{\rm r}}_{100}$  & $\tau^{z_{\rm r}}_{100}$ & $\zeta^{z_{\rm r}}_{50}$  & $\tau^{z_{\rm r}}_{50}$ &\\
\hline
7.5         & $6.71$ & $0.68$ & $6.74$ & $0.76$ & $6.71$ & $0.83$ & $6.60$ & $0.75$ & $6.53$ & $0.85$  & \\
9.0         & $7.98$ & $1.29$ & $7.96$ & $1.22$ & $7.93$ & $1.32$ & $7.75$ & $1.72$ &  $7.67$ & $1.37$ & \\
10.5        & $9.49$ & $2.00$ & $9.50$ & $2.12$ & $9.33$ & $2.03$ & $8.99$ & $1.71$ & $8.94$ & $2.03$  & \\
12.0        & $11.10$ & $2.58$ & $11.33$ & $3.05$ & $10.98$ & $2.90$ & $10.39$ & $2.38$ & $10.40$ & $2.84$ &\\
13.5        & $12.36$ & $2.35$ & $13.12$ & $3.63$ & $12.75$ & $3.72$ & $11.93$ & $3.07$ & $12.02$ & $3.69$ &\\
15.0        & $13.27$ & $1.77$ & $14.36$ & $3.26$ & $14.64$ & $4.49$ & $13.66$ & $3.75$ & $13.86$ & $4.60$ &\\
\hline
\end{tabular}
\end{center}
\end{table*}

\subsection{Probability density function}
\label{Sec:Appendix:PDF}
\cite{Miralda:2000} provided a convenient four-parameter fit to the volume-weighted probability density function (PDF) of the gas density 
at redshifts $z = 2, 3$ and $4$ obtained from the L10 hydrodynamical simulation described in \cite{Miralda:1996}. In addition,
they provided a prescription for extrapolating this fit to higher redshifts. Here we compare the volume-weighted PDF obtained from our 
default simulation (\textit{r9L6N256}) at $z = 6$ to their fit. Because we find that their set of fitting 
parameters provides a somewhat poor description of this PDF, we also provide an updated set of parameters that yields a fit which more accurately
describes the one-point distribution of gas densities at $z = 6$ in our default simulation.
\par
\begin{figure}
  \includegraphics[width=0.49\textwidth]{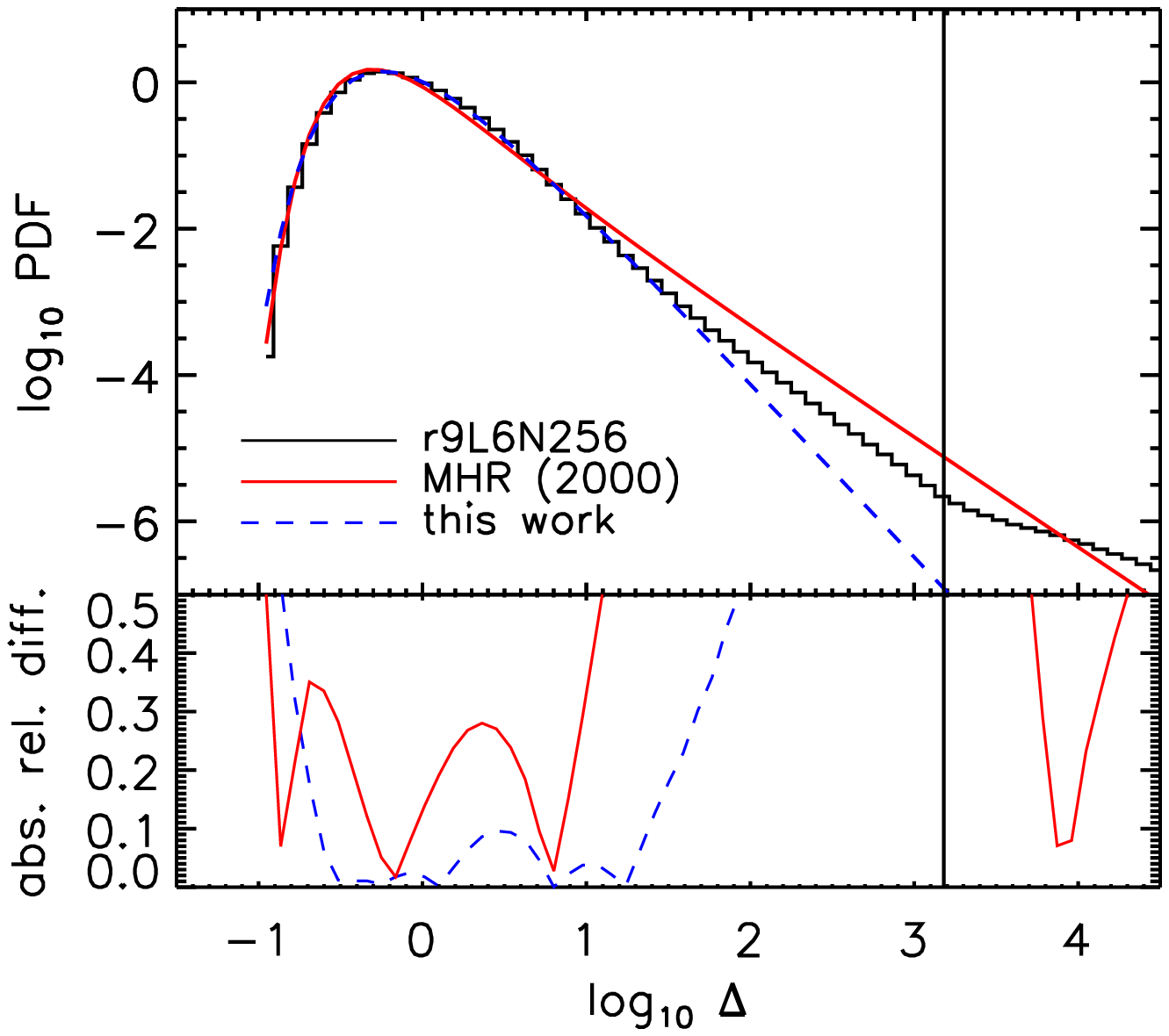}
  \caption{Top panel: Volume-weighted PDF of the baryon overdensity $\Delta$ (per unit $\log_{10} \Delta$) 
    at $z = 6$ obtained from our default simulation \textit{r9L6N256} (black solid histogram).  
    For comparison, the red solid line shows the four-parameter fit given by Eq.~\ref{Eq:PDF:Fit}, with the parameter values 
    taken from \protect\cite{Miralda:2000}. The blue dashed line shows our best fit of Eq.~\ref{Eq:PDF:Fit} to the PDF.
    Bottom panel: Absolute value of the relative differences of the fits with respect to the PDF. 
    In both panels, the vertical line indicates the overdensity corresponding to the onset of star formation.
}

  \label{Fig:PDF:Fit}
\end{figure}

In Fig.~\ref{Fig:PDF:Fit} we show the volume-weighted PDF of the gas density per unit $\log_{10} \Delta$. 
We have investigated the convergence of this PDF with respect to the resolution and the size of the simulation box, using the same set of simulations
that we employed to study the convergence of the clumping factor in Section~\ref{Sec:Convergence}. We find that the PDF is converged with respect 
to changes in the resolution. It is converged with respect to changes in the size of the simulation box for overdensities $\log_{10} \Delta \lesssim 2$.
For larger overdensities, the values of the PDF obtained from our default simulation may not yet be fully converged. 
We note that at $z = 6$ the PDFs obtained from the simulations \textit{r[$z_{\rm r}$]L6N256z} that employ a reheating redshift $z_{\rm r} > 9$
are almost undistinguishable from that obtained from our default simulation.
\par
We compare the PDF to the fit given by \cite{Miralda:2000},
\begin{equation}
\mathcal{P}_{\rm V} (\Delta) d\Delta = A \exp \left [ - \frac{(\Delta^{-2/3} - C_0)^2}{2(2\delta_0/3)^2}\right] \Delta^{-\beta} d\Delta,
\label{Eq:PDF:Fit}
\end{equation}
using their values for the parameters\footnote{Note that the value for A given in \cite{Miralda:2000} is too large by a factor of $\ln 10$.} 
$A = 0.375, \delta_0 = 1.09, \beta = 2.50$ and $C_0 = 0.880$. As can be seen from Fig.~\ref{Fig:PDF:Fit}, 
for overdensities $1 \lesssim  \log_{10} \Delta \lesssim 2$ the PDF obtained from our default simulation is significantly steeper than predicted by the fit. 
Using the simulations \textit{r9L6N256W3} and \textit{r9L6N256W1}, we verified 
that the steepness of the slope of the PDF over this range of overdensities is sensitive to the cosmological parameters employed. 
Note that the PDF does also not asymptote to $\mathcal{P}_{\rm V} (\Delta) \propto \Delta^{-\beta}$, as predicted by Eq.~\ref{Eq:PDF:Fit}. 
For overdensities larger than the overdensity for the onset of star formation, the PDF is instead governed by the effective equation of state 
characteristic for the multi-phase star-forming gas.
\par
We fitted Eq.~\ref{Eq:PDF:Fit} to our PDF over the range $-1 \le \log_{10} \Delta \le 2$, constraining the parameters $A$ and $C_0$ by the requirement that the
total integral over the volume- and mass-weighted PDF must be normalized to unity. This yields values 
$A = 3.038, \delta_0 = 1.477, \beta = 3.380$ and $C_0 = -0.932$. Note that the fit based on these parameter values 
still does not provide a good description of the PDF for overdensities $\log_{10} \Delta \gtrsim 2$. 

\bsp

\label{lastpage}


\begin{thebibliography}{99} 

\bibitem[\protect\citeauthoryear{Abel \& Haehnelt}{1999}]{Abel:1999} 
Abel T., Haehnelt M.~G., 1999, ApJ, 520, L13 

\bibitem[\protect\citeauthoryear{Barkana \& Loeb}{1999}]{Barkana:1999} 
Barkana R., Loeb A., 1999, ApJ, 523, 54 

\bibitem[\protect\citeauthoryear{Barkana \& Loeb}{2001}]{Barkana:2001} 
Barkana R., Loeb A., 2001, PhR, 349, 125 

\bibitem[\protect\citeauthoryear{Barkana \& Loeb}{2004}]{Barkana:2004} 
Barkana R., Loeb A., 2004, ApJ, 609, 474 

\bibitem[\protect\citeauthoryear{Benson et al.}{2001}]{Benson:2001} 
Benson A.~J., Nusser A., Sugiyama N., Lacey C.~G., 2001, MNRAS, 320, 153 

\bibitem[\protect\citeauthoryear{Bolton, Meiksin, \& White}{2004}]{Bolton:2004} 
Bolton J., Meiksin A., White M., 2004, MNRAS, 348, L43 

\bibitem[\protect\citeauthoryear{Bolton \& Haehnelt}{2007}]{Bolton:2007} 
Bolton J.~S., Haehnelt M.~G., 2007, MNRAS, 382, 325 

\bibitem[\protect\citeauthoryear{Bouwens et al.}{2004}]{Bouwens:2004} 
Bouwens R.~J., et al., 2004, ApJ, 606, L25 

\bibitem[\protect\citeauthoryear{Bouwens et al.}{2006}]{Bouwens:2006} 
Bouwens R.~J., Illingworth G.~D., Blakeslee J.~P., Franx M., 2006, ApJ, 653, 53 

\bibitem[\protect\citeauthoryear{Bouwens et al.}{2007}]{Bouwens:2007} 
Bouwens R.~J., Illingworth G.~D., Franx M., Ford H., 2007, ApJ, 670, 928 

\bibitem[\protect\citeauthoryear{Bouwens et al.}{2008}]{Bouwens:2008} 
Bouwens R.~J., Illingworth G.~D., Franx M., Ford H., 2008, preprint (arXiv:0803.0548) 

\bibitem[\protect\citeauthoryear{Bruzual \& Charlot}{2003}]{Bruzual:2003}
Bruzual G., Charlot S., 2003, MNRAS, 344, 1000 

\bibitem[\protect\citeauthoryear{Bunker et al.}{2004}]{Bunker:2004} 
Bunker A.~J., Stanway E.~R., Ellis R.~S., McMahon R.~G., 2004, MNRAS, 355, 374 

 \bibitem[\protect\citeauthoryear{Chabrier}{2003}]{Chabrier:2003} 
Chabrier G., 2003, PASP, 115, 763 

\bibitem[\protect\citeauthoryear{Chiu, Fan, \& Ostriker}{2003}]{Chiu:2003} 
Chiu W.~A., Fan X., Ostriker J.~P., 2003, ApJ, 599, 759 

\bibitem[\protect\citeauthoryear{Ciardi et al.}{2006}]{Ciardi:2006} 
Ciardi B., Scannapieco E., Stoehr F., Ferrara A., Iliev I.~T., Shapiro P.~R., 2006, MNRAS, 366, 689 

\bibitem[\protect\citeauthoryear{Ciardi \& Salvaterra}{2007}]{Ciardi:2007} 
Ciardi B., Salvaterra R., 2007, MNRAS, 381, 1137 

\bibitem[\protect\citeauthoryear{Collin-Souffrin}{1991}]{Collin:1991} 
Collin-Souffrin S., 1991, A\&A, 243, 5 

\bibitem[\protect\citeauthoryear{Chuzhoy, Kuhlen, \& Shapiro}{2007}]{Chuzhoy:2007} 
Chuzhoy L., Kuhlen M., Shapiro P.~R., 2007, ApJ, 665, L85

\bibitem[\protect\citeauthoryear{Crain et al.}{2007}]{Crain:2007} 
Crain R.~A., Eke V.~R., Frenk C.~S., Jenkins A., McCarthy I.~G., Navarro 
J.~F., Pearce F.~R., 2007, MNRAS, 377, 41 

\bibitem[\protect\citeauthoryear{Dalla Vecchia \& Schaye}{2008}]{caius:2008}
Dalla Vecchia C., Schaye J., 2008, MNRAS, 387, 1431 

\bibitem[\protect\citeauthoryear{Dijkstra et al.}{2004}]{Dijkstra:2004} 
Dijkstra M., Haiman Z., Rees M.~J., Weinberg D.~H., 2004, ApJ, 601, 666 

\bibitem[\protect\citeauthoryear{Efstathiou}{1992}]{Efstathiou:1992} 
Efstathiou G., 1992, MNRAS, 256, 43P 

\bibitem[\protect\citeauthoryear{Fan, Carilli,\& Keating}{2006}]{Fan:2006} 
Fan X., Carilli C.~L., Keating B., 2006, ARA\&A, 44, 415 

\bibitem[\protect\citeauthoryear{Faucher-Gigu{\`e}re et al.}{2008}]{Faucher:2008} 
Faucher-Gigu{\`e}re C.-A., Lidz A., Hernquist L., Zaldarriaga M., 2008, ApJ, 688, 85 

\bibitem[\protect\citeauthoryear{Ferland et al.}{1998}]{Ferland:1998} 
Ferland G.~J., Korista K.~T., Verner D.~A., Ferguson J.~W., Kingdon J.~B., Verner E.~M., 1998, PASP, 110, 761 

\bibitem[\protect\citeauthoryear{Furlanetto \& Oh}{2005}]{Furlanetto:2005} 
Furlanetto S.~R., Oh S.~P., 2005, MNRAS, 363, 1031 

\bibitem[\protect\citeauthoryear{Furlanetto, Oh, \& Briggs}{2006}]{Furlanetto:2006} 
Furlanetto S.~R., Oh S.~P., Briggs F.~H., 2006, PhR, 433, 181 

\bibitem[\protect\citeauthoryear{Furlanetto, Haiman, \& Oh}{2008}]{Furlanetto:2008} 
Furlanetto S.~R., Haiman Z., Oh S.~P., 2008, ApJ, 686, 25 

\bibitem[\protect\citeauthoryear{Glover \& Brand}{2003}]{Glover:2003} 
Glover S.~C.~O., Brand P.~W.~J.~L., 2003, MNRAS, 340, 210 

\bibitem[\protect\citeauthoryear{Glover}{2007}]{Glover:2007} 
Glover S.~C.~O., 2007, MNRAS, 379, 1352 

\bibitem[\protect\citeauthoryear{Gnedin \& Ostriker}{1997}]{Gnedin:1997} 
Gnedin N.~Y., Ostriker J.~P., 1997, ApJ, 486, 581 

\bibitem[\protect\citeauthoryear{Gnedin \& Hui}{1998}]{Gnedin:1998} 
Gnedin N.~Y., Hui L., 1998, MNRAS, 296, 44 

\bibitem[\protect\citeauthoryear{Gnedin}{2000a}]{Gnedin:2000a} 
Gnedin N.~Y., 2000a, ApJ, 535, 530 

\bibitem[\protect\citeauthoryear{Gnedin}{2000b}]{Gnedin:2000b} 
Gnedin N.~Y., 2000b, ApJ, 542, 535 

\bibitem[\protect\citeauthoryear{Gnedin, Kravtsov, \& Chen}{2008a}]{Gnedin:2008a} 
Gnedin N.~Y., Kravtsov A.~V., Chen H.-W., 2008, ApJ, 672, 765 

\bibitem[\protect\citeauthoryear{Gnedin}{2008b}]{Gnedin:2008b} 
Gnedin N.~Y., 2008, ApJ, 673, L1 

\bibitem[\protect\citeauthoryear{Haardt \& Madau}{2001}]{Haardt:2001} 
Haardt F., Madau P., 2001, in the proceedings of XXXVI Rencontres de Moriond, preprint (astroph/0106018)

\bibitem[\protect\citeauthoryear{Haiman, Rees, \& Loeb}{1997}]{Haiman:1997} 
Haiman Z., Rees M.~J., Loeb A., 1997, ApJ, 476, 458 

\bibitem[\protect\citeauthoryear{Haiman, Abel, \& Madau}{2001}]{Haiman:2001} 
Haiman Z., Abel T., Madau P., 2001, ApJ, 551, 599 

\bibitem[\protect\citeauthoryear{Hoeft et al.}{2006}]{Hoeft:2006} 
Hoeft M., Yepes G., Gottl{\"o}ber S., Springel V., 2006, MNRAS, 371, 401 

\bibitem[\protect\citeauthoryear{Hui \& Haiman}{2003}]{Hui:2003} 
Hui L., Haiman Z., 2003, ApJ, 596, 9 

\bibitem[\protect\citeauthoryear{Iliev, Scannapieco, \& Shapiro}{2005a}]{Iliev:2005a} 
Iliev I.~T., Scannapieco E., Shapiro P.~R., 2005a, ApJ, 624, 491 

\bibitem[\protect\citeauthoryear{Iliev, Shapiro, \& Raga}{2005b}]{Iliev:2005b} 
Iliev I.~T., Shapiro P.~R., Raga A.~C., 2005b, MNRAS, 361, 405 

\bibitem[\protect\citeauthoryear{Iliev et al.}{2007}]{Iliev:2007} 
Iliev I.~T., Mellema G., Shapiro P.~R., Pen U.-L., 2007, MNRAS, 376, 534 

\bibitem[\protect\citeauthoryear{Inoue, Iwata, \& Deharveng}{2006}]{Inoue:2006} 
Inoue A.~K., Iwata I., Deharveng J.-M., 2006, MNRAS, 371, L1 

\bibitem[\protect\citeauthoryear{Kennicutt}{1998}]{Kennicutt:1998} 
Kennicutt R.~C., Jr., 1998, ApJ, 498, 541 

\bibitem[\protect\citeauthoryear{Kitayama \& Ikeuchi}{2000}]{Kitayama:2000} 
Kitayama T., Ikeuchi S., 2000, ApJ, 529, 615 

\bibitem[\protect\citeauthoryear{Kohler, Gnedin, \& Hamilton}{2007}]{Kohler:2007} 
Kohler K., Gnedin N.~Y., Hamilton A.~J.~S., 2007, ApJ, 657, 15 

\bibitem[\protect\citeauthoryear{Komatsu et al.}{2008}]{Komatsu:2008} 
Komatsu E., et al., 2008, preprint (arXiv:0803.0547) 

\bibitem[\protect\citeauthoryear{Kuhlen \& Madau}{2005}]{Kuhlen:2005} 
Kuhlen M., Madau P., 2005, MNRAS, 363, 1069 

\bibitem[\protect\citeauthoryear{Lacey \& Cole}{1994}]{Lacey:1994}
Lacey C., Cole S., 1994, MNRAS, 271, 676

\bibitem[\protect\citeauthoryear{Lehnert \& Bremer}{2003}]{Lehnert:2003} 
Lehnert M.~D., Bremer M., 2003, ApJ, 593, 630 

\bibitem[\protect\citeauthoryear{Machacek, Bryan, \& Abel}{2003}]{Machacek:2003} 
Machacek M.~E., Bryan G.~L., Abel T., 2003, MNRAS, 338, 273 

\bibitem[\protect\citeauthoryear{Madau \& Efstathiou}{1999}]{Madau:1999a} 
Madau P., Efstathiou G., 1999, ApJ, 517, L9 

\bibitem[\protect\citeauthoryear{Madau, Haardt, \& Rees}{1999}]{Madau:1999} 
Madau P., Haardt F., Rees M.~J., 1999, ApJ, 514, 648 

\bibitem[\protect\citeauthoryear{Madau}{2000}]{Madau:2000} 
Madau P., 2000, RSPTA, 358, 2021 

\bibitem[\protect\citeauthoryear{Madau et al.}{2004}]{Madau:2004} 
Madau P., Rees M.~J., Volonteri M., Haardt F., Oh S.~P., 2004, ApJ, 604, 484 

\bibitem[\protect\citeauthoryear{Maio et al.}{2007}]{Maio:2007} 
Maio U., Dolag K., Ciardi B., Tornatore L., 2007, MNRAS, 379, 963 

\bibitem[\protect\citeauthoryear{Malhotra et al.}{2005}]{Malhotra:2005} 
Malhotra S., et al., 2005, ApJ, 626, 666 

\bibitem[\protect\citeauthoryear{Mannucci et al.}{2007}]{Mannucci:2007} 
Mannucci F., Buttery H., Maiolino R., Marconi A., Pozzetti L., 2007, A\&A, 461, 423

\bibitem[\protect\citeauthoryear{Mesinger \& Dijkstra}{2008}]{Mesinger:2008} 
Mesinger A., Dijkstra M., 2008, preprint (arXiv:0806.3090) 

\bibitem[\protect\citeauthoryear{Miralda-Escud{\'e} \& Rees}{1994}]{Miralda:1994} 
Miralda-Escud{\'e} J., Rees M.~J., 1994, MNRAS, 266, 343 

\bibitem[\protect\citeauthoryear{Miralda-Escud{\'e} et al.}{1996}]{Miralda:1996} 
Miralda-Escud{\'e} J., Cen R., Ostriker J.~P., Rauch M., 1996, ApJ, 471, 582 

\bibitem[\protect\citeauthoryear{Miralda-Escud{\'e}, Haehnelt, \& Rees}{2000}]{Miralda:2000} 
Miralda-Escud{\'e} J., Haehnelt M., Rees M.~J., 2000, ApJ, 530, 1 

\bibitem[\protect\citeauthoryear{Miralda-Escud{\'e}}{2003}]{Miralda:2003} 
Miralda-Escud{\'e} J., 2003, ApJ, 597, 66 

\bibitem[\protect\citeauthoryear{Navarro \& Steinmetz}{1997}]{Navarro:1997} 
Navarro J.~F., Steinmetz M., 1997, ApJ, 478, 13 

\bibitem[\protect\citeauthoryear{Oesch et al.}{2008}]{Oesch:2008} 
Oesch P.~A., et al., 2008, preprint (arXiv:0804.4874)

\bibitem[\protect\citeauthoryear{Oh}{2001}]{Oh:2001} 
Oh S.~P., 2001, ApJ, 553, 499

\bibitem[\protect\citeauthoryear{Oh \& Haiman}{2003}]{Oh:2003} 
Oh S.~P., Haiman Z., 2003, MNRAS, 346, 456 

\bibitem[\protect\citeauthoryear{Okamoto, Gao, \& Theuns}{2008}]{Okamoto:2008} 
Okamoto T., Gao L., Theuns T., 2008, preprint (arXiv:0806.0378) 

\bibitem[\protect\citeauthoryear{Padmanabhan}{1993}]{Padmanabhan:1993} 
Padmanabhan T., Structure Formation in the Universe, Cambridge University Press (1993) 

\bibitem[\protect\citeauthoryear{Panagia et al.}{2005}]{Panagia:2005} 
Panagia N., Fall S.~M., Mobasher B., Dickinson M., Ferguson H.~C., Giavalisco M., Stern D., Wiklind T., 2005, ApJ, 633, L1 

\bibitem[\protect\citeauthoryear{Pawlik \& Schaye}{2008}]{Pawlik:2008} 
Pawlik A.~H., Schaye J., 2008, submitted, preprint (arXiv:0812.2913) 

\bibitem[\protect\citeauthoryear{Razoumov \& Sommer-Larsen}{2006}]{Razoumov:2006} 
Razoumov A.~O., Sommer-Larsen J., 2006, ApJ, 651, L89 

\bibitem[\protect\citeauthoryear{Ricotti \& Ostriker}{2004}]{Ricotti:2004} 
Ricotti M., Ostriker J.~P., 2004, MNRAS, 352, 547 

\bibitem[\protect\citeauthoryear{Ryu et al.}{1993}]{Ryu:1993} 
Ryu D., Ostriker J.~P., Kang H., Cen R., 1993, ApJ, 414, 1 

\bibitem[\protect\citeauthoryear{Sawicki \& Thompson}{2006}]{Sawicki:2006} 
Sawicki M., Thompson D., 2006, ApJ, 648, 299 

\bibitem[\protect\citeauthoryear{Schaye}{2001}]{Schaye:2001}
Schaye J., 2001, ApJ, 559, 507 

\bibitem[\protect\citeauthoryear{Schaye}{2004}]{Schaye:2004} 
Schaye J., 2004, ApJ, 609, 667 

\bibitem[\protect\citeauthoryear{Schaye \& Dalla Vecchia}{2008}]{Schaye:2008} 
Schaye J., Dalla Vecchia C., 2008, MNRAS, 383, 1210 

\bibitem[\protect\citeauthoryear{Seljak \& Zaldarriaga}{1996}]{Seljak:1996} 
Seljak U., Zaldarriaga M., 1996, ApJ, 469, 437

\bibitem[\protect\citeauthoryear{Shapiro, Giroux, \& Babul}{1994}]{Shapiro:1994} 
Shapiro P.~R., Giroux M.~L., Babul A., 1994, ApJ, 427, 25 

\bibitem[\protect\citeauthoryear{Shapiro, Iliev, \& Raga}{2004}]{Shapiro:2004} 
Shapiro P.~R., Iliev I.~T., Raga A.~C., 2004, MNRAS, 348, 753 

\bibitem[\protect\citeauthoryear{Shapley et al.}{2003}]{Shapley:2003} 
Shapley A.~E., Steidel C.~C., Pettini M., Adelberger K.~L., 2003, ApJ, 588, 65 

\bibitem[\protect\citeauthoryear{Spergel et al.}{2007}]{Spergel:2007}
Spergel D.~N., et al., 2007, ApJS, 170, 377

\bibitem[\protect\citeauthoryear{Spergel et al.}{2003}]{Spergel:2003}
Spergel D.~N., et al., 2003, ApJS, 148, 175

\bibitem[\protect\citeauthoryear{Springel \& Hernquist}{2003}]{Springel:2003} 
Springel V., Hernquist L., 2003, MNRAS, 339, 289 

\bibitem[\protect\citeauthoryear{Springel}{2005}]{Springel:2005} 
Springel V., 2005, MNRAS, 364, 1105 

\bibitem[\protect\citeauthoryear{Srbinovsky \& Wyithe}{2007}]{Srbinovsky:2007} 
Srbinovsky J.~A., Wyithe J.~S.~B., 2007, MNRAS, 374, 627 

\bibitem[\protect\citeauthoryear{Stanway, Bunker, \& McMahon}{2003}]{Stanway:2003} 
Stanway E.~R., Bunker A.~J., McMahon R.~G., 2003, MNRAS, 342, 439 

\bibitem[\protect\citeauthoryear{Stiavelli, Fall, \& Panagia}{2004}]{Stiavelli:2004} 
Stiavelli M., Fall S.~M., Panagia N., 2004, ApJ, 610, L1 

\bibitem[\protect\citeauthoryear{Susa \& Umemura}{2004}]{Susa:2004} 
Susa H., Umemura M., 2004, ApJ, 600, 1 

\bibitem[\protect\citeauthoryear{Theuns et al.}{2002}]{Theuns:2002} 
Theuns T., Schaye J., Zaroubi S., Kim T.-S., Tzanavaris P., Carswell B., 2002, ApJ, 567, L103 

\bibitem[\protect\citeauthoryear{Thoul \& Weinberg}{1996}]{Thoul:1996} 
Thoul A.~A., Weinberg D.~H., 1996, ApJ, 465, 608 

\bibitem[\protect\citeauthoryear{Tittley \& Meiksin}{2007}]{Tittley:2007} 
Tittley E.~R., Meiksin A., 2007, MNRAS, 380, 1369 

\bibitem[\protect\citeauthoryear{Valageas \& Silk}{1999}]{Valageas:1999} 
Valageas P., Silk J., 1999, A\&A, 347, 1 

\bibitem[\protect\citeauthoryear{Veilleux, Cecil, \& Bland-Hawthorn}{2005}]{Veilleux:2005} 
Veilleux S., Cecil G., Bland-Hawthorn J., 2005, ARA\&A, 43, 769 

\bibitem[\protect\citeauthoryear{Venkatesan, Giroux, \& Shull}{2001}]{Venkatesan:2001} 
Venkatesan A., Giroux M.~L., Shull J.~M., 2001, ApJ, 563, 1 

\bibitem[\protect\citeauthoryear{Wiersma, Schaye \& Smith}{2008}]{Wiersma:2008} 
Wiersma R.~P.~C., Schaye J., Smith B.~D., 2008, MNRAS, in press, preprint (arXiv:0807.3748)

\bibitem[\protect\citeauthoryear{Wise \& Abel}{2005}]{Wise:2005} 
Wise J.~H., Abel T., 2005, ApJ, 629, 615 

\bibitem[\protect\citeauthoryear{Wise \& Abel}{2008}]{Wise:2008} 
Wise J.~H., Abel T., 2008, ApJ, 684, 1 

\bibitem[\protect\citeauthoryear{Yan \& Windhorst}{2004}]{Yan:2004} 
Yan H., Windhorst R.~A., 2004, ApJ, 600, L1 

\bibitem[\protect\citeauthoryear{Y{\"u}ksel et al.}{2008}]{Yuksel:2008} 
Y{\"u}ksel H., Kistler M.~D., Beacom J.~F., Hopkins A.~M., 2008, ApJ, 683, L5 

\bibitem[\protect\citeauthoryear{Zuo \& Phinney}{1993}]{Zuo:1993} 
Zuo L., Phinney E.~S., 1993, ApJ, 418, 28 


\end{thebibliography}
\end{document}